\documentclass[useAMS,usenatbib]{mn2e}

\usepackage{graphicx}
\usepackage{aas_macros}
\usepackage{placeins}

\usepackage{graphicx}    

\title[Connecting the dots]{Connecting the dots: A versatile model for the atmospheres of tidally locked Super-Earths}
\author[L. Carone, R. Keppens and L. Decin]{L. Carone$^{1}$\thanks{E-mail:
ludmila.carone@wis.kuleuven.be (LC)},  R.
Keppens$^{1}$, and L. Decin$^{2}$\\
$^{1}$Centre for mathematical Plasma Astrophysics, Department of Mathematics, KU Leuven, Celestijnenlaan 200B, 3001 Leuven, Belgium\\
$^{2}$Instituut voor Sterrenkunde, KU Leuven, Celestijnenlaan 200D, 3001 Leuven, Belgium}
\begin{document}

\date{22.7.2014}

\pagerange{\pageref{firstpage}--\pageref{lastpage}} \pubyear{accepted}

\maketitle

\label{firstpage}

\begin{abstract}
Radiative equilibrium temperatures are calculated for the troposphere of a tidally locked Super-Earth based on a simple greenhouse model, using Solar System data as a guideline. These temperatures provide in combination with a Newtonian relaxation scheme thermal forcing for a 3D atmosphere model using the dynamical core of the Massachusetts Institute of Technology global circulation model (MITgcm). Our model is of the same conceptional simplicity than the model of \citet{HeldSuarez1994} and is thus computationally fast. Furthermore, because of the coherent, general derivation of radiative equilibrium temperatures, our model is easily adaptable for different planets and atmospheric scenarios. As a case study relevant for Super-Earths, we investigate a Gl581g-like planet with Earth-like atmosphere and irradiation and present results for two representative rotation periods of $P_{rot}=10$~days and $P_{rot}=36.5$~days. Our results provide proof of concept and highlight interesting dynamical features for the rotating regime $3<P_{rot}<100$~days, which was shown by \cite{Edson2011} to be an intermediate regime between equatorial superrotation and divergence.
We confirm that the $P_{rot}=10$~days case is more dominated by equatorial superrotation dynamics than the $P_{rot}=36.5$~days case, which shows diminishing influence of standing Rossby-Kelvin waves and increasing influence of divergence at the top of the atmosphere. We argue that this dynamical regime change relates to the increase in Rossby deformation radius, in agreement with previous studies. However, we also pay attention to other features that are not or only in partial agreement with other studies, like, e.g., the number of circulation cells and their strength, the role and extent of thermal inversion layers, and the details of heat transport.
\end{abstract}

\begin{keywords}
planets and satellites: atmospheres --planets and satellites: terrestrial planets -- methods: numerical.
\end{keywords}


\section{Introduction}
\label{Introduction}
There are many roads towards modelling and understanding of planetary atmospheres. Dry three dimensional global circulation models (3D GCMs) with idealized thermal forcing are one possible method to gain insights into atmospheric large-scale dynamics and are even to this date relevant to the Earth climate community. In a landmark paper, \cite{HeldSuarez1994} (\textit{HS94}) reduced the complexity of a 3D GCM to the very basics by replacing the detailed radiative forcing, turbulence and moist convection descriptions with very simple forcing prescriptions in a full 3D hydrodynamical core. Dry 3D GCMs have the advantage that they can easily adjust to varying planetary and atmosphere parameters and that they are computationally efficient. This not only allows to compare the dynamical cores of different climate models with each other, which was the initial intention for the \textit{HS94} Earth benchmark, but also allows to disentangle non-linear interdependencies between forcing parameters and large-scale circulation, which is much more difficult in more complex models \citep{Held2005}. The \textit{HS94} benchmark has been used to study, to name only a few examples, the effect of temperature forcing on timescales of variability in the extratropical atmosphere \citep{Gerber2007}, sensitivity of extratropical circulation to tropopause height that might change due to global warming \citep{Lorenz2007}, and winds in the upper equatorial troposphere \citep{Kraucunas2005}. Despite the simplifications, the main characteristics of the Earth atmosphere dynamics already emerge from \textit{HS94} like Hadley, Ferrel and polar circulation cells and jet streams (see, e.g., \citet{Bordi2009}). Identification of uplifting circulation branches allows further to identify regions of cloud formation that are particularly habitable.

Ultimately, complex models are needed that take into account complex atmosphere chemistry (e.g. \citet{Grenfell2013}, \citet{Hu2012} among others), radiative transfer (e.g. \cite{Kataria2014}) and hydrological cycles \citep{Menou2013}. These models aim to derive spectra (e.g. \cite{Bailey2012}, \cite{Hedelt2013}) for comparison with observational data. Their results all critically depend on the dynamically established thermodynamic equilibrium. Therefore the following questions should be addressed: Is the resolution sufficient? Are the numerical representation at the poles and the atmosphere boundaries (surface, top) adequate, and more importantly, are all relevant physical processes accounted for? Keeping in mind that already for the Earth it is difficult to parametrize sub-grid processes, like eddies, clouds, planetary boundary layer, wave drag etc., it is widely understood that there is even less certainties for exoplanets, where even the composition of their atmospheres and surface is unknown. The appeal of dry idealized GCMs in the context of exoplanets is thus immediately clear: They are ideally suited to provide benchmarking.

We show in this paper that it is indeed possible to construct a simplified model suitable for all terrestrial planets with the same level of complexity as \textit{HS94}, if one examines rigorously the underlying assumptions of the temperature forcing given in \textit{HS94}. Using a simplified greenhouse model and the skin layer assumption, we developed a new temperature forcing scheme that is suitable for a tidally locked planet with negligible obliquity, that can also be used for planets with different atmospheric composition. We propose, therefore, a versatile and computationally efficient terrestrial planet atmosphere model that retains full control over parameters and physical prescriptions - in particular subgrid processes - by incorporating empirical knowledge about Solar System planetary atmospheres.

We aim to compute models of the atmospheres of habitable extrasolar planets that ultimately will be helpful for observations, making it clear that tidally-locked Super-Earths in the habitable zones of late K and M-dwarf stars will be our first targets. This is because such planets will be easier to detect than truly Earth-size planets due to a more favourable planet/star contrast ratio. Tidal interaction will then synchronize the planet's rotation with its revolution leading to rotation periods between 8 and 100 days as pointed out by \citet{Edson2011}.

Indeed, the rotation period $P_{rot}$ is one key parameter in shaping the dynamics of exoplanet atmospheres. The atmospheres of fast rotating tidally locked Earth-like planets ($P_{rot}=1$~day) are according to \citet{Merlis2010} dominated by waves and eddies with a westerly jet at the equator, whereas for slow rotation ($P_{rot}=360$~days), it is dominated by divergent circulation. \cite{Edson2011} found the phase state transition between the fast and slow rotating regime between $4 \to 5$~days for dry planets and $3 \to 4$~days for aquaplanets. Another transition was found for a rotation period of 100~days. Interestingly, \cite{Edson2011} consistently report two circulation cells for each hemisphere even for $P_{rot}=100$~days, whereas most other earlier studies - including \cite{Joshi1997} and \cite{Joshi2003} - report just one.

In the following, we present our model in Section~\ref{sec: model}. We present two representative simulations with $P_{rot}=10$ and $P_{rot}=36.5$~days for the intermediate rotation regime $3\leq P_{rot}\leq 100$~days, identified by \cite{Edson2011}, and discuss our results in Section~\ref{sec: results}. We show that our model yields results that are generally consistent with previous studies using more complex GCMs. In particular, we discuss: superrotation and divergent flow, cyclonic vortices, circulation and temperature distribution in context with possible habitability of tidally locked terrestrial planets. We also highlight differences with published work and postulate possible reasons for their deviation. We conclude with a short summary of our results in Section~\ref{sec: Conclusion} and give an outlook on future work. We emphasize that the atmospheric dynamics is surprisingly complex in the intermediate rotation regime ($ 3 \leq P_{rot} \leq 100$~days) due to the mixture between divergent and superrotating circulation that warrants a dedicated parameter study. We highlight the potential of our model to perform such a parameter study and outline further possible investigations to test basic model assumptions like surface friction and thermal forcing variation, in particular, at the nightside.

\section{The model}
\label{sec: model}

In the following, we describe the specific set-up of the model for terrestrial exoplanets with a greenhouse atmosphere. We stress once again that our goal is to construct a conceptionally simple model that yields the large scale dynamics in the troposphere with maximum computational efficiency.

\subsection{The dynamical core: MITgcm}
\label{sec: dynamical core}

The dynamical core of the MITgcm developed at MIT\footnote{http://mitgcm.org} \citep{Adcroft2004} uses the finite-volume method to solve the primitive hydrostatical equations (HPE) that write as horizontal momentum
\begin{equation}
\frac{D\vec{v}}{Dt} +f\vec{k} \times \vec{v}=-\nabla_p \Phi +\mathcal{\vec{F}}_v,\label{eq: mom}
\end{equation}
vertical stratification
\begin{equation}
\frac{\partial \Phi}{\partial p}=-\frac{1}{\rho},
\end{equation}
continuity
\begin{equation}
\vec{\nabla}_p\cdot \vec{v}+\frac{\partial \omega}{\partial p}=0,
\end{equation}
equation of state for an ideal gas
\begin{equation}
p=\rho \left(R/\mu\right)T,
\end{equation}
and thermal forcing equation
\begin{equation}
\frac{D\theta}{Dt}=\frac{\theta}{c_pT}\mathcal{F}_\theta,\label{eq:thermo}
\end{equation}
where the total derivative is
\begin{equation}
\frac{D}{Dt}=\frac{\partial}{\partial t}+\vec{v}\cdot \nabla_p+\omega \frac{\partial}{\partial p}.
\end{equation}
In these equations, $\vec{v}$ is the horizontal velocity and $\omega=Dp/Dt$ is the vertical velocity component in an isobaric coordinate system, where $p$ is pressure and $t$ is time. $\Phi=gz$ is the geopotential, with $z$ being the height of the pressure surface, $g$ the surface gravity, and $f=2\Omega \sin \nu$ is the Coriolis parameter, with $\Omega$ being the planet's angular velocity and $\nu$ the latitude at a given location. $\vec{k}$ is the local vertical unit vector, $T$, $\rho$ and $c_p$ are the temperature, density, and specific heat at constant pressure. $\theta=T\left(p/p_s\right)^\kappa$ is the potential temperature with $\kappa$ being the ratio of the specific gas constant $R/\mu$ to $c_p$, where  $R=8.314$~JK$^{-1}$mol$^{-1}$ is the ideal gas constant and $\mu$ is the molecular mean mass of the atmosphere, and $p_s$ is the surface pressure. $\mathcal{F}_{\theta,v}$ are the horizontal and thermal forcing terms, respectively. Note that, in this case, $\mathcal{\vec{F}}_v$ is a vector and $\mathcal{F}_\theta$ is a scalar.

The \textit{HPEs} are solved in $\eta$ vertical coordinates on a staggered Arakawa C grid using a curvilinear cubed-sphere horizontal coordinate system \citep{Showman2009,Marshall1997}. $\eta$ is in the absence of topography defined as
\begin{equation}
\eta=\frac{p-p_t}{p_s-p_t},
\end{equation}
where $p_t$ denotes the pressure at the top of the model atmosphere.

The quasi-second order Adams-Bashforth time-stepping scheme with a stabilizing parameter $\epsilon_{AB}=0.1$ was used for all explicit terms in the momentum equations (see, e.g., \citet{Durran1991} for a discussion of time-stepping methods for the numerical simulation of advection in the atmosphere). The model top at $p_t=0$ has vanishing vertical velocity ($\omega=0$) as boundary condition. Pressure at the surface is treated in the implicit free surface form of the pressure equation \citep{Marshall1998}, which allows to fluctuate the surface pressure around the constant $p_s$.

For our study, we aim to investigate the dynamics in the troposphere, the `weather region', that comprises the majority of the atmosphere's mass. Therefore, we need to resolve this region with sufficient vertical spacing. \citet{HengVogt2011} used a linear vertical spacing with 20 levels seperated at 500 mbar with a surface pressure of $p_s=1$~bar in accordance to \textit{HS94}. Indeed, it was confirmed by \citet{Will1998} that this vertical resolution is sufficient to resolve the general structure of the Earth's troposphere. Therefore, we also use it for our study. We adopt the C32 cubed-sphere grid for the horizontal resolution that was tested successfully with \textit{HS94} by \citet{Marshall2004} and is also used by \citet{Zalucha2013}. The sphere is subdivided into six tiles, each with $32\times 32$ elements, thus there are $32\times 32 \times 20$ volume elements per tile. This horizontal gridding corresponds to a global resolution in longitude-latitude of $128\times 64$ or $2.^{\circ}8\times 2.^{\circ}8$ and translates to a mean horizontal resolution of $\Delta x,y\approx 41$~km for a planet with $1.45$~Earth radii.

\begin{figure}
\includegraphics[width=0.495\textwidth]{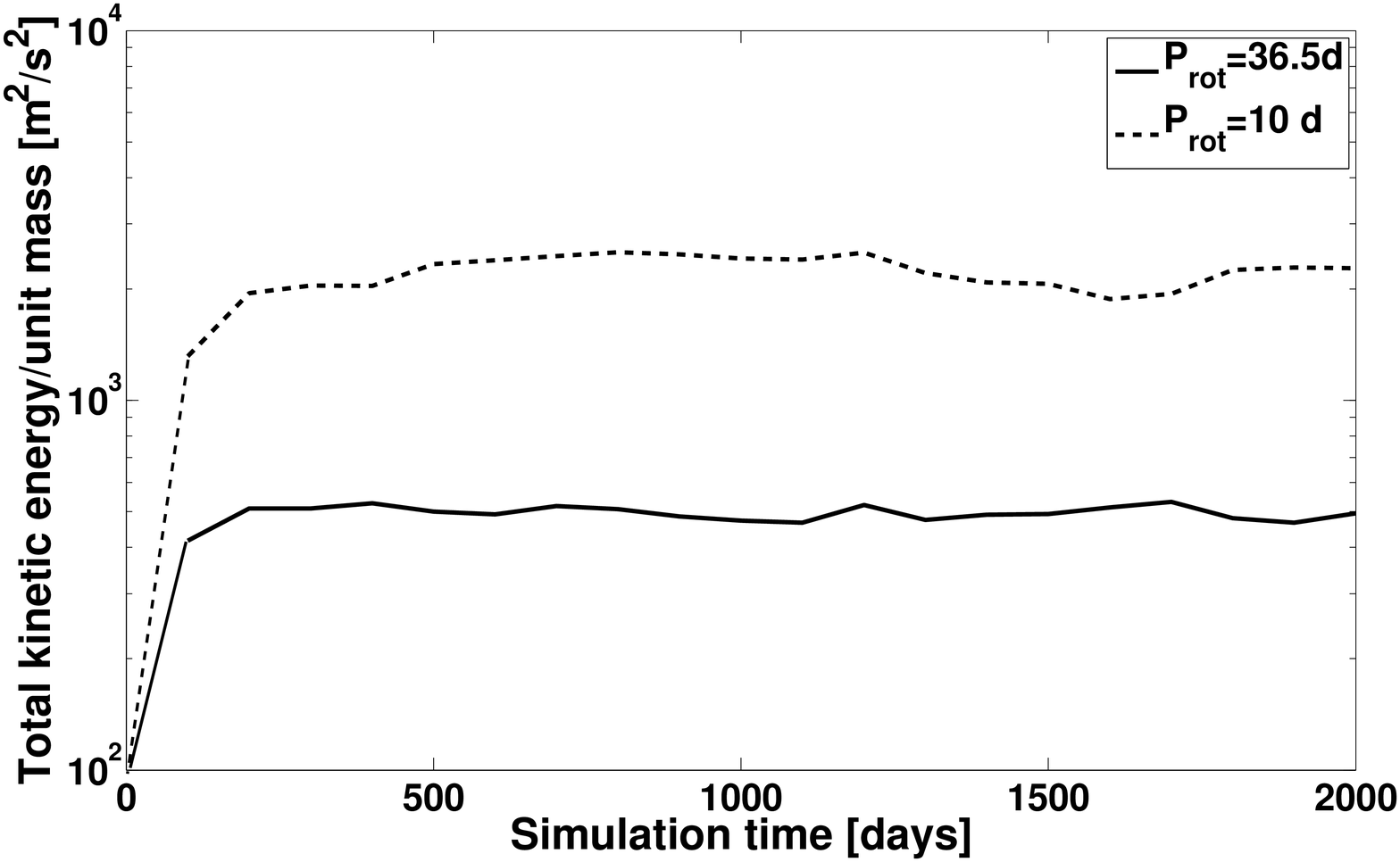}
\caption{Kinetic energy per unit mass for our GCM runs after initialization at $t=0$~days. The kinetic energy increases steeply and reaches, in any case, equilibrium after 400 days.}
\label{fig:KE}
\end{figure}

\citet{HengVogt2011} assumed that the spin-up of the atmospheric system from initial conditions at constant temperature $T=264$~K would be concluded after $t_{init}=200$~Earth~days. Figure~\ref{fig:KE} shows the globally averaged kinetic energy per unit mass for our two runs with $P_{rot}=10$ and $36.5$~days, where we found that our simulations are just barely at equilibrium after 200~days. Therefore, we have set $t_{init}=400$~Earth~days to be on the safe side. Data generated before $t_{init}$ were discarded; the simulation was run subsequently for $t_{run}=1000$~Earth~days and averaged over this time period. Time averaging over 1000 days should filter out small scale variations and transient waves, allowing to bring forth the basic stable large scale dynamics.

\subsection{Friction timescale}
\label{sec: friction}
We prescribe Rayleigh friction between the atmosphere and the surface via

\begin{equation}
\mathcal{\vec{F}}_v=-\frac{1}{\tau_{fric}}\vec{v},
\end{equation}
where $\tau_{fric}$ is defined as
\begin{equation}
\tau_{fric}=\tau_{s,fric} \max\left(0,\frac{\frac{p}{p_s}-\frac{p_{PBL}}{p_s}}{1-\frac{p_{PBL}}{p_s}}\right),
\end{equation}
where $p_{PBL}$ is the pressure at the upper limit of the planetary boundary layer (PBL) and $\tau_{s,fric}$ is the maximum surface friction. Both values are somewhat unconstrained even for terrestrial planets in the Solar System. \textit{HS94} use $\tau_{s,fric}=1$~day and $p_{PBL}=700$~mbar for the Earth. For Mars, on the other hand, the possible values for $\tau_{s,fric}$ range between $0.1-30$~days (see Table~\ref{tab: PBL}). We assume for now $\tau_{s,fric}=1$~day and $p_{PBL}=700$~mbar like in \textit{HS94}.

\begin{table}
\caption{Friction timescales and planetary boundary layer extent (PBL) for some Solar System planets}
\begin{tabular}{l|c|p{1.8cm}|p{3cm}|}
\hline
planet & $\tau_{s,fric}$ & $p_{PBL}$ & Reference\\
 & [days] & &\\
\hline
Earth & 1 & $0.7\times p_s$ & \citet{HeldSuarez1994}\\
Venus & 25 & $0.87\times p_s$ & \citet{Lee2007}\\
Mars & 10-27.2, 0.2 & $(0.91-0.78)\times p_s$ & \citet{Haberle1997}, \citet{Joshi1995}\\
\hline
\end{tabular}
\label{tab: PBL}
\end{table}

Like \cite{Zalucha2013}, we prescribe an upper sponge-layer with Rayleigh friction. This sponge layer prevents non-physical wave reflection at the upper boundary but can also be physically justified by assuming gravity wave braking at high pressure levels. We implemented the Rayleigh dampening profile suggested by \citet{Polvani2002} by adding to $\mathcal{\vec{F}}_v$ the term $-k_R\vec{v}$, where $k_R$ is:
\begin{eqnarray}
k_R&=&k_{max}\left[\frac{p_{sponge}-p}{p_{sponge}}\right]^2 \quad \mathrm{if}~ p\geq p_{sponge}\\
k_R&=&0 \qquad\mathrm{if}~ p< p_{sponge},\nonumber
\end{eqnarray}
with $p_{sponge}=0.1$~bar and $k_{max}=1/80$~days$^{-1}$ appropriate for damping in the middle stratosphere \citep{Jablo2011}.

\subsection{Temperature forcing timescale}

The Newtonian relaxation scheme is applied through an external temperature forcing
\begin{equation}
\mathcal{F}_T=\frac{T_{eq}(\nu,\phi,p)-T}{\tau_{rad}},\label{eq: F_T}
\end{equation}
where the temperature $T$ in the model atmosphere is driven with the relaxation timescale $\tau_{rad}$ towards an equilibrium temperature $T_{eq}$ that will be defined in the following section. $\mathcal{F}_T$ can easily be transformed into $\mathcal{F}_\theta$, by using the definition of the potential temperature $\theta=T\left(p/p_s\right)^\kappa$. The relaxation timescale is estimated to first order by (e.g., \citet{Showman2002}):
\begin{equation}
\tau_{rad}= \frac{c_p p_s}{4 g \sigma T_s^3}\label{eq:tau_rad},
\end{equation}
where $T_s$ is the mean surface temperature, $p_s$ is the surface pressure and $\sigma$ is the Stefan-Boltzmann constant. This estimate yields realistic values for low pressures $p\leq 1$ bar (see Table~\ref{tab: tau_T}). For higher pressures ($p \geq 10$~bars), however, comparison with Venus between the computed radiative timescales from more detailed radiative transfer calculations and the estimate given above ($\tau_{rad}\approx 100$~years instead of 112~days stemming from the first order estimate) and comparison between the first order estimate and the timescales calculated in \citet{Showman2008} show that equation~(\ref{eq:tau_rad}) may be an underestimate by orders of magnitude.

For atmospheric models of dense atmospheres ($p_s>>1$~bar), therefore, either larger $\tau_{rad}$ should be investigated or instead of assuming a single $\tau_{rad}$ for the whole vertical atmosphere column, equation~(6) of \cite{Showman2008} should be used instead to calculate $\tau_{rad}$ for a given vertical layer with pressure $p_0$ and temperature $T_0$, following the steps outlined in that work. In this study, however, we aim to build a \textit{HS94}-like model of a tidally locked terrestrial exoplanet with an Earth-like troposphere with $p_s= 1$~bar covering a limited pressure range between 1000~mbar to 100~mbar, for which the first order estimate is valid and sufficient.

\begin{table}
\caption{Temperature forcing timescales for some  planets}
\begin{tabular}{l|p{1.3cm}|p{0.8cm}|p{3cm}|}
\hline
planet & $\tau_{rad}$ (radiative transfer) & $\tau_{rad}$ \newline(first order) & Reference\\
\hline
 & [days] & [days] &\\
\hline
HD~209458b & 2 & 1.1 & \citet{Showman2002},\citet{Iro2005}\\
Earth & 40 & 20 & \citet{HeldSuarez1994}\\
Venus & 115-19000 & 112 & \citet{Pollack1975}\\
Mars & 2 & 0.6 & \citet{Haberle1997}\\
\hline
\end{tabular}
\label{tab: tau_T}
\end{table}

In addition, we implement the dry convection scheme  of \citet{Molteni2002} to treat static instabilities at the substellar point. It diffuses vertically dry static energy $s=c_pT+\Phi$, if the static stability is locally violated. That is the case, if $d\theta/dp>0$, where we have assumed a vertical diffusion timescale of $\tau_{vds}=1$~day.

\subsection{Equilibrium temperature $T_{eq}$}

The equilibrium temperature $T_{eq}(\nu,\phi,p)$, which is used in the thermal forcing $\mathcal{F}_T$ (equation~(\ref{eq: F_T})), generally depends on the planet's latitude $\nu$, longitude $\phi$ and pressure level $p$ and is prescribed as follows:

\begin{equation}
T_{eq}(\nu,\phi,p)=\max\left(T_{NS}(p),T_{DS}(\nu,\phi,p)\right),\label{eq: T_eq}
\end{equation}
where $T_{NS}$ are nightside and $T_{DS}$ are dayside temperatures. The dayside thermal forcing is allowed to drop to very low temperatures at the poles and the terminator, until the nightside temperature conditions are met. Furthermore, we assume that the temperature drops with decreasing pressure (increasing height) following the dry adiabat at the dayside and a nitrogen condensing moist adiabat on the nightside, until the atmosphere reaches the tropopause temperature $T_{tp}$. The temperature remains further on at $T_{tp}$ and constitutes a vertically isothermal layer on top of our troposphere. 

The lower boundary of the radiative-convective temperature profile on the dayside is derived by calculating the surface temperature using a first order greenhouse model adopting an optical depth at the surface, $\tau_s$, appropriate for an atmosphere of Earth-like composition. On the nightside, we set the lower boundary by assuming as surface temperature the nitrogen condensation temperature for $p_s=1$~bar.

To calculate the upper boundary, we adopt for both, day and nightside, the skin layer concept by \citet{Pierrehumbert} to determine the top of the convective layer, that is, the tropopause: First, we assume very small but non-zero values for optical depth and emissivity: $\tau_{tp}\approx 0$ and $\epsilon_{tp}\approx 0$. 
\begin{figure}
\includegraphics[width=0.48\textwidth]{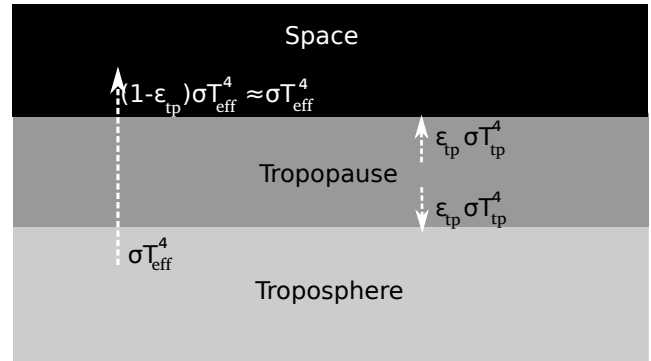}
\caption{Visualization of the skin layer concept used to derive the tropopause temperature. An optically very thin atmosphere layer of very small absorptance and emittance $\epsilon_{tp}\approx 0$ is overlaid on top of a troposphere that is heated by the black body flux of the atmosphere, $\sigma T_{eff}^4$. The skin layer's contribution to the overall heat balance, however, is negligible.}
\label{fig: skin}
\end{figure}
It is further assumed that the skin layer's density is too small to transport heat efficiently into and from the layer by convection.

In this case, little of the upwelling infrared radiation is absorbed in the skinlayer due to its small absorptance $\epsilon_{tp}$ and it emits only a small amount of infrared back radiation towards the lower atmosphere, $\epsilon_{tp} \sigma T^4_{tp}$. Therefore, the skin layer is only heated by infrared upwelling of the atmosphere with negligible contribution to the overall heat budget and the energy balance for the tropopause under skin layer assumptions reads (Figure~\ref{fig: skin} and see, e.g. \citet{Pierrehumbert}):
\begin{equation}
2\epsilon_{tp} \sigma T^4_{tp}\approx \epsilon_{tp}\sigma T^4_{eff}(\nu,\phi),
\end{equation}
and yields for the skin-layer temperature that defines the tropopause:
\begin{equation}
T_{tp}=0.5^{1/4}T_{eff}(\nu,\phi) \label{eq: tp_temp},
\end{equation}
where $T_{eff}(\nu,\phi)$ is the effective or black body temperature of the atmosphere that will be calculated in the following for the day and nightside, respectively.

\subsubsection{Condensing nightside}

In our nominal model, it is assumed that the temperatures at the nightside may plummet so low that they reach the condensation point for liquid nitrogen. For now, we prescribe a `Martian pole scenario' \citep{Toon1980}: It is assumed that the vertical temperature relaxes towards the saturated moist adiabat of the main ingredient of the atmosphere, $N_2$, that is described in the limit of a one-component condensable atmosphere as:
\begin{equation}
T_{*}(p)=\frac{T_{ref}}{1-\frac{\left(R/\mu\right)T_{ref}}{L}\ln\frac{p}{p_{ref}}}.
\end{equation}
$T_{ref}$ and $p_{ref}$ are the vaporization reference temperature and pressure, which is $T_{ref}=77$~K at $p_{ref}=1$~bar for nitrogen, respectively, $L=1.98\times 10^5$~J/kg is the vaporization latent heat release \citep{Pierrehumbert}, $R$ is the ideal gas constant, $\mu=28$~g/mol is the mean molecular mass of $N_2$. Furthermore, we assume for now that surface pressure variation due to condensation is negligible. It should be noted that \cite{Zalucha2013} used a similar prescription for the nightside of their tidally locked $H_2O$ atmosphere planet.

The vertically isothermal tropopause on top of the troposphere is calculated by assuming that a nitrogen condensing atmosphere does not contain any greenhouse gases so that for the optical depth at the surface: $\tau_s\approx 0$. In this case, the surface temperature $T_*(p_s)$ is equivalent to the effective temperature. Therefore, equation (\ref{eq: tp_temp}) can be applied  to derive the tropopause temperature on the nightside with $T_*(p_s)=T_{eff}(\nu,\phi)$ as:
\begin{equation}
T_{tp}=0.5^{1/4}T_{*}(p_s).
\end{equation}

The total nightside equilibrium temperature description used in equation~(\ref{eq: T_eq}) is accordingly:
\begin{equation}
T_{NS}(p)=\max\left(T_{tp},T_{*}(p)\right).
\end{equation}

\cite{Pierrehumbert} states that in a situation where $\tau_s\approx 0$, the skin layer would extend down to the ground and the atmosphere assume $T_{tp}$, if radiative heating would be the only heat transfer mechanism. However, because the surface temperature is warmer than an atmosphere at $T_{tp}$, convection would kick in. Convection would then establish a vertical gradient along the dry adiabat for a non-condensing atmosphere. We assume that the same holds equivalently for condensing atmospheres with the difference that the atmosphere relaxes along the above described saturated moist adiabat. 

If the atmosphere is condensing nitrogen on the nightside in some vertical layers and not in others, then only the condensing layers will follow the moist adiabat, where we neglect for now transport of latent heat to the neighbouring non-condensing cells. However, as will be shown later, this is not an issue with the two simulations shown in this work, because the nightside temperatures are at all pressure levels well above the nitrogen condensation limit due to the efficient dynamic heat transport from the dayside. We may revise this prescription in future work, in particular, when going to lower surface pressures. Furthermore, we will investigate for specific cases if it's possible to maintain the atmosphere in a partly or fully condensing situation without total atmospheric collapse.

\subsubsection{Illuminated dayside temperature with zero obliquity}

For the dayside, it is assumed that the incident flux $F$ is distributed with the cosine of the zenith angle $\cos \zeta$ across the sphere, which is for zero obliquity and a tidally locked planet (see also \cite{Joshi1997}) geometrically related to the latitude and longitude by:
\begin{equation}
\cos \zeta=\cos\phi\cos\nu.
\end{equation}
Using Stefan Boltzmann's law $F=\sigma T^4$, the horizontal temperature distribution has to be $\propto \cos^{1/4}\phi\cos^{1/4}\nu$. This yields for the dayside effective temperatures consequently:

\begin{equation}
T_{eff}(\nu,\phi)=T_{eff,max}\cos^{1/4}\nu \cos^{1/4}\phi,\label{eq: eff_ds}
\end{equation}
with the maximum effective planetary temperature:
\begin{equation}
T_{eff,max}=\left[\frac{I_0(1-\alpha)}{\sigma}\right]^{1/4}.\label{eq: eff_ds_max}
\end{equation}
The incoming flux is $F=I_0(1-\alpha)$, where $\alpha$ is the albedo of the planet and the incident stellar flux $I_0$ of a star with luminosity $L_*$, radius $R_*$ and effective temperature $T_{eff,*}$ on a planet at distance $a$ is
\begin{equation}
I_0=\sigma T_{eff,*}^4\left(\frac{R_*}{a}\right)^2.
\end{equation}
From the analytic solution of the two-stream approximation for a greenhouse atmosphere, the following relation for the dayside surface temperature follows, where $\tau_s$ is the optical depth at the surface and the value of $\gamma$ depends on which approximation is used to treat scattering in the atmosphere:
\begin{equation}
T_{DS,s}(\nu,\phi)=(\gamma\tau_s+1)^{1/4} T_{eff}(\nu,\phi).\label{eq: opaque}
\end{equation}
\cite{Zalucha2013} used in their equation~(8) the Eddington approximation for which $\gamma=0.75$. In this work, we use the hemi-isotropic or hemispheric mean approximation for which $\gamma=1$ (see, e.g., \citet{Toon1989,Pierrehumbert,Heng2014} for a discussion of different scattering treatments in the two-stream formalism). We argue here that light from M dwarf stars, that are the relevant host stars for tidally locked planets with Earth-like thermal forcing, is less subject to Rayleigh scattering and that therefore $\gamma=1$ is here more appropriate \citep{Toon1989}. However, it is easy to change this assumption for other type of host stars and the difference in temperature between the two approximations is small\footnote{\textbf{It is $\approx 7$~K for $\tau_s=0.62$ and $T_{eff,P}=255$~K appropriate for Earth taken from Table~\ref{tab: SS_vert}.}}.

For an optically thin atmosphere (optical depth at surface $\tau_s<1$), we may also formulate the surface dayside temperatures in terms of the leaky greenhouse model (see, e.g. \citet{Marshall}):
\begin{equation}
T_{DS,s}(\nu,\phi)=\left(\frac{2}{2-\epsilon}\right)^{1/4} T_{eff}(\nu,\phi),\label{eq: thin}
\end{equation}
where the emissivity $\epsilon$ is connected with the surface optical depth $\tau_s$ via
\begin{equation}
\frac{2}{2-\epsilon}-1=\gamma\tau_s. \label{eq: tau}
\end{equation}

In any case, the skin layer concept is applied again to define the upper boundary of the convective troposphere as a vertically isothermal tropopause for dayside conditions:
\begin{equation}
T_{tp}=0.5^{1/4}T_{eff,max}\cos^{1/4}\nu \cos^{1/4}\phi,
\end{equation}
combining equations~(\ref{eq: tp_temp}) and (\ref{eq: eff_ds}).

Assuming now that the vertical temperature gradient in the troposphere follows the dry adiabatic index $dT/dz=-\Gamma_d=-g/c_p$ because the troposphere is subjected to convection, we have the following prescription for the equilibrium temperature:
\begin{eqnarray}
&&T_{eq}(\nu,\phi,p)=\max\{T_{NS}(p),\nonumber\\
&&\max \left[T_{tp},T_{DS,s}(\nu,\phi)\left(\frac{p}{p_s}\right)^{R/c_p}\right]\}\label{eq:horiz_DS}.
\end{eqnarray}
In Table~\ref{tab: SS_vert}, we collect albedo, solar irradiance and calculate the effective temperature and tropopause temperature for selected Solar System bodies. From the mean global surface temperature $T_s$, assuming distribution of solar energy over the whole sphere, $\tau_s$ can be inferred via equations~(\ref{eq: opaque}) and~(\ref{eq: thin}). The derived $\tau_s$-values serve us as `ground truths'. We also introduce here the scale height, which is defined as $H=R T_s/\mu g$.

Figure~\ref{fig:Earth_forcing} shows the vertical equilibrium temperature profile for an Earth-like atmosphere ($\epsilon =0.8$, $\alpha=0.3$, and $p_s=1$~bar) on a tidally-locked planet subjected to Earth-like stellar irradiation. At both hemispheres, the temperature drops continuously from the surface temperature to colder temperatures with increasing height, following a dry adiabat at the dayside and the condensation profile at the nightside. Consequently, the effective temperature $T_{eff}$ and the tropopause temperature $T_{tp}$ are automatically reached at some pressure level. For our calculations, the tropopause is located at approximately 450~mbar at the dayside and 200~mbar at the nightside. It has to be noted that an increase of optical depth at the dayside -- and thus to higher surface temperatures than computed here -- would lead to a shift of the effective temperature and tropopause location to higher altitudes, that is, to lower pressure levels. A decrease of optical depth would lead correspondingly to a downward shift in height. This connection between optical depth, temperature and altitude explains why $T_{eff}$ is already reached at surface pressure at the nightside, where $\tau_s\approx 0$ is assumed. The nightside tropopause of the prescribed equilibrium temperature $T_{eff}$ is, however, located at higher altitude than the dayside tropopause because the vertical temperature gradient due to condensation is much less steep than the vertical temperature gradient due to convection. Note, however, that dynamics will change this picture. It tends to decrease the vertical temperature gradient due to convection (Table~\ref{tab: SS_vert}) and, as will be shown later, to increase the vertical gradient on the nightside as the temperatures don't drop low enough to allow for condensation. Thus, the actual tropopause is shifted to higher altitudes at the dayside, whereas for the nightside the raising of surface temperature and increase in vertical temperature gradient appear to compensate for each other.

\begin{table*}
\caption{Selected Solar System atmosphere data}
\begin{tabular}{l|p{2cm}|p{2cm}|p{2cm}|p{2cm}|p{2cm}|}
\hline
& Earth & Venus & Mars & Titan & Jupiter\\
\hline
$I_0$ [Wm$^{-2}$]$^{1}$ & 1368$^{3}$ & 2614$^{3}$ & 589$^{3}$& 15& 50.5$^{3}$\\
albedo$^{1}$ & 0.3$^{3}$ & 0.77 & 0.25$^3$&0.22$^{3}$& 0.34$^{3}$\\
$T_{eff,P}$ [K]$^2$& 255 & 210 & 210 & 85& 124.4$^b$\\
$T_{tp}$ [K]$^2$&214 & 177 & 177& 71.5 & 105$^b$\\
$\gamma$ $^1$& 7/5& 9/7& 9/7 &7/5&7/5\\
constituent $^{1,3}$& N$_2$ & CO$_2$&CO$_2$ & N$_2$ & H$_2$\\
$\mu$[g/mol]$^d$ &28 & 44 & 44& 28&2\\
$c_p$ [J/gK]$^{c,2}$& 1.04 & 0.85 & 0.85 & 1.04 & 14.55\\
$H$ [km]$^{2}$ &8.7 & 15.6 & 11.1 &20.7$^f$ &28\\
$g$[ms$^{-2}$]$^{1,3}$&9.8 &8.9&3.7& 1.35& 24.8$^a$\\
$p_s$[bar]$^{1,3}$& 1&92&0.006&1.47& 1$^a$\\
$T_s$[K]$^2$ &288&737&220& 94 &167$^a$\\
$\epsilon$~$^{e}$ & 0.8 & - & 0.33 & 0.66 & -\\
$\tau_s$~$^{e}$ & 0.62&110&0.2&0.5&2.25\\
$-\frac{dT}{dz}_{dry}$\newline$\left[\frac{K}{km}\right]^2$ & 9.4&10.4&4.4 &1.3 &1.7\\
$-\frac{dT}{dz}_{real}$\newline$\left[\frac{K}{km}\right]^1$ & 3-9.4&7.7&1.6-4&0.56-1.24&1.7\\
literature &\cite{Marshall}, \cite{Pierrehumbert}&\cite{Irvine1968}, \cite{Kliore1980}, \cite{Lewis1971}, \cite{Pierrehumbert}&\cite{Goody1972}, \cite{Pierrehumbert}&\cite{McKay1997}, \cite{McKay1999}, \cite{Elachi2005}, \cite{Brown2010}, \cite{Pierrehumbert} & \cite{Lindal1981}, \cite{Hanel1981}, \cite{Pierrehumbert}\\
\hline
\end{tabular}
\label{tab: SS_vert}
\newline
$a$) lower boundary of troposphere set at $p$=1 bar,
$b$) including 60\% contribution from internal heat flux\newline
$c$) $\kappa= R/(\mu c_p)=1-1/\gamma$,
$d$) equal $\mu$ of main constituent\newline
$e$) calculated using equations~(\ref{eq: thin}), (\ref{eq: tau}), and (\ref{eq: opaque}), setting $T_s=T_{DS,s}$.\newline
$f$) Due to haze in the upper atmosphere, the real value is $H=40$~km \citep{Justus2004}\newline
$1$) literature, $2$) calculated using equation~(\ref{eq: tp_temp}), 3)http://nssdc.gsfc.nasa.gov/planetary/factsheet/\newline
\end{table*}

\subsection{Optical depth and albedo}

We assume that the optical depth is linearly dependent on pressure (e.g., \cite{Zalucha2013}) and that therefore the optical depth decreases from its maximal value at the surface $\tau_s$ to the tropopause $\tau_{tp}\approx 0$. Furthermore, non-condensing nitrogen dominated atmospheres with surface pressures in the order of $1$~bar are expected to be optically thin (Table~\ref{tab: SS_vert}). It is thus justified to assume an optically thin, leaky greenhouse model for the calculation of the equilibrium temperature $T_{eq}$ at the dayside in this work. The effect of a varied content of greenhouse gases can then, in principle, be investigated by varying the emissivity $\epsilon$ and thus the optical depth $\tau_s$ between approximately zero and one\footnote{ To keep the skin layer concept, $\tau_s$ can never be exactly zero and furthermore $\tau_{tp}<\tau_s$ has to hold. Indeed, this case was assumed for the derivation of the equilibrium temperature at the nightside in the limit of nitrogen condensation.}. $\tau_s \approx 0$ corresponds then to an atmosphere with negligible greenhouse gas and $\tau_s=1$ to an atmosphere with large greenhouse gas content, like the $H_2O$ atmosphere used by \citet{Zalucha2013} that has indeed $\tau_s=1$ for $p_s=1$~bar.

Another important factor that determines the thermal forcing on a tidally locked planet is the reflectivity of clouds, ice coverage, soil that are encapsulated in the albedo coefficient $\alpha$, which may also be varied between zero and one. In this work, we want to demonstrate that our model is indeed suitable for the investigation of atmospheric dynamics of terrestrial tidally-locked planets. Thus, we focus for now on Earth-like atmospheres to compare our results with previous studies and assume thus $p_s=1000$~mbar, an albedo of $\alpha=0.3$, and $\epsilon =0.8$ for the dayside as for the Earth (Table~\ref{tab: SS_vert}), and $\epsilon\approx 0$ and $\tau_s\approx 0$ at the nightside because the temperature is allowed to drop so low that all greenhouses gases freeze out. We note, however, that it is easy to change our model here to accommodate atmospheres with different optical depths and albedo.

\subsection{Investigated parameter space: Tidally locked Super-Earths with rotation periods in between $3~\leq P_{rot}\leq 100$~days}

We have constructed a simplified versatile atmosphere model for a tidally locked terrestrial planet with the same elegant simplicity as the \textit{HS94} benchmark, without carrying over assumptions that are only valid for the Earth. This is the drawback of the model of \citet{HengVogt2011} that retains some Earth-centric features.

\begin{figure}
\includegraphics[width=0.495\textwidth]{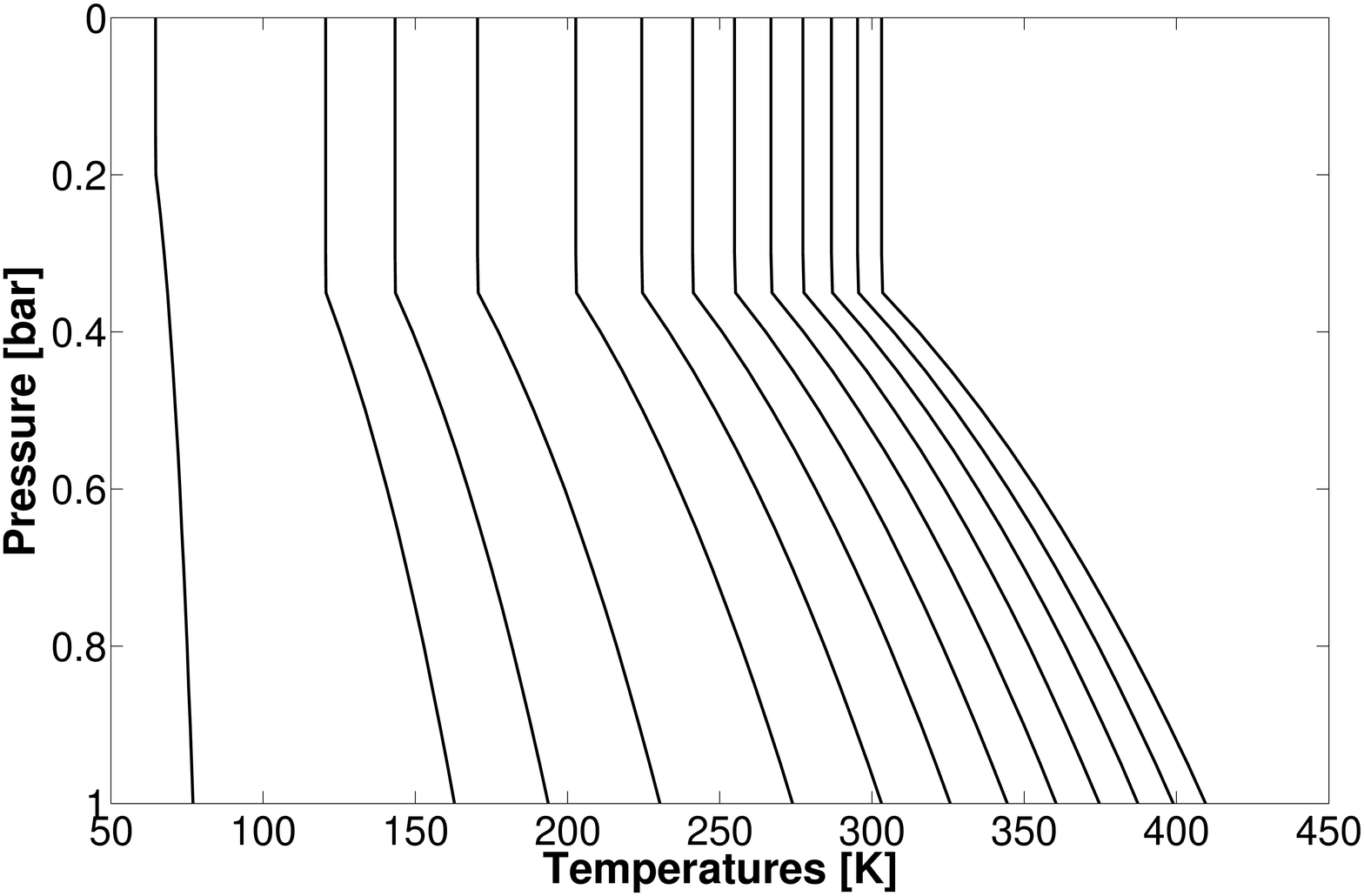}
\caption{Vertical equilibrium temperature profiles for Earth-like irradiation ($I_0=1368$~W/m$^2$ and $\alpha=0.3$) at the nightside/terminator region (coldest profile), i.e., $\cos \zeta =0$ and for $\cos \zeta=0.025, 0.05$ and $\cos \zeta$ between 0.1 and 1 in steps of 0.1.}
\label{fig:Earth_forcing}
\end{figure}

Many previous studies focus on planets around M dwarf stars as these planets are the most numerous main sequence stars and also the ones for which terrestrial exoplanets are more readily detected and observed. They broadly fall into two categories: complex Earth models that are used for tidally locked planets but are constrained to the thermal forcing regime of Earth (e.g, \cite{Edson2011}, \cite{Joshi1997}, \cite{Merlis2010}). On the other side, there are less complex models that allow to get out of the comfort zones of the Earth models in terms of forcing and composition, with more or less sophistication in their radiative transfer schemes. E.g., \cite{Zalucha2013} use a gray two stream formalism with Eddington approximation and \citet{Kataria2014} use the more complex SPARC radiative transfer code.

We use, like \cite{Zalucha2013}, a gray two stream formalism but with the hemi-isotropic approximation for scattering for the surface temperature and calculate a radiative-convective equilibrium temperature between the surface temperature and an upper boundary temperature. Furthermore, we focus on the investigation of the dynamics in the troposphere and cap our upper atmosphere with a vertically isothermal tropopause, using the skin layer concept to identify the top of the convective troposphere. We will show that our set-up is sufficient to resolve many dynamical features reported by other more complex and/or extended models but with the major benefit that we are very flexible with our parametrization.

In this work, we focus on Earth composition atmospheres, as most dynamical studies have focused on this type of terrestrial planets to address the prospect of habitability. We assume the parameters reported by \cite{HengVogt2011} as typical for a Super-Earth: $R_P=1.45 R_{Earth}$ and $M_P=3.1 M_{Earth}$, resulting in a surface gravity of $g=14.3$~m/s$^2$. We furthermore use $P_{rot}=36.5$~days, like in the nominal model of \cite{HengVogt2011}, allowing to intercompare our results directly. We further investigate $P_{rot}=10$~days to bridge the gap to the faster rotating regime.

The radiative timescale $\tau_{rad}$ is calculated using equation~(\ref{eq:tau_rad}) and we use Earth-like composition and irradiation to set the equilibrium temperature profile $T_{eq}(p,\nu,\phi)$. It should be noted that we also allow the nightside to relax towards the condensation temperature which is different from the procedure prescribed by \cite{Zalucha2013} who allowed to cool the nightside without bound and instantaneously reset the temperatures to the condensation vertical profile if the temperature dropped below the $H_2O$ condensation temperature. \citet{Joshi1997} and \citet{Edson2011}, on the other hand, imposed no forcing on the nightside and allowed the nightside to cool without bound.

Table~\ref{tab: Planetary parameters} lists the planetary and atmospheric parameters used in our simulations.

\begin{table}
\caption{Main Super-Earth planetary and atmospheric parameters for our two reference models}
\begin{tabular}{|l|c|}
Parameter & \\
\hline
Planetary radius $R_p$  & $1.45 R_{Earth}$\\
Planetary mass $M_P$ & $3.1 M_{Earth}$\\
Surface gravity $g$& 14.3~m/s$^2$ \\
Obliquity & $0^\circ$\\
$c_p$ & 1.04 J/gK\\
$p_s$ & 1000 mbar \\
main constituent & $N_2$\\
molecular mass $\mu$ & 28 g/mol\\
atmospheric emissivity $\epsilon$ & 0.8\\
surface optical depth $\tau_s$ & 0.62 \\
adiabatic index $\gamma$ & 7/5 \\
Surface friction timescale $\tau_{fric}$ & 1 day\\
Radiative timescale at substellar point $\tau_{rad}$ & 13 days\\
Radiative timescale at nightside $\tau_{rad}$ & 813 days\\
Rotation period $P_{rot}$ & 10~days, 36.5~days\\
\hline
\end{tabular}
\label{tab: Planetary parameters}
\end{table}

\section{Results and Discussion}
\label{sec: results}

In the following, we will discuss flow patterns maintained as a consequence of the imposed forcing averaged over 1000~days after the model reached steady state, that is, after $t_{init}=400$~days as is shown in Figure~\ref{fig:KE} and discussed in Section~\ref{sec: dynamical core}.

\subsection{Horizontal flow}
\label{sec: Horiz}

\begin{figure*}
\includegraphics[width=0.95\textwidth]{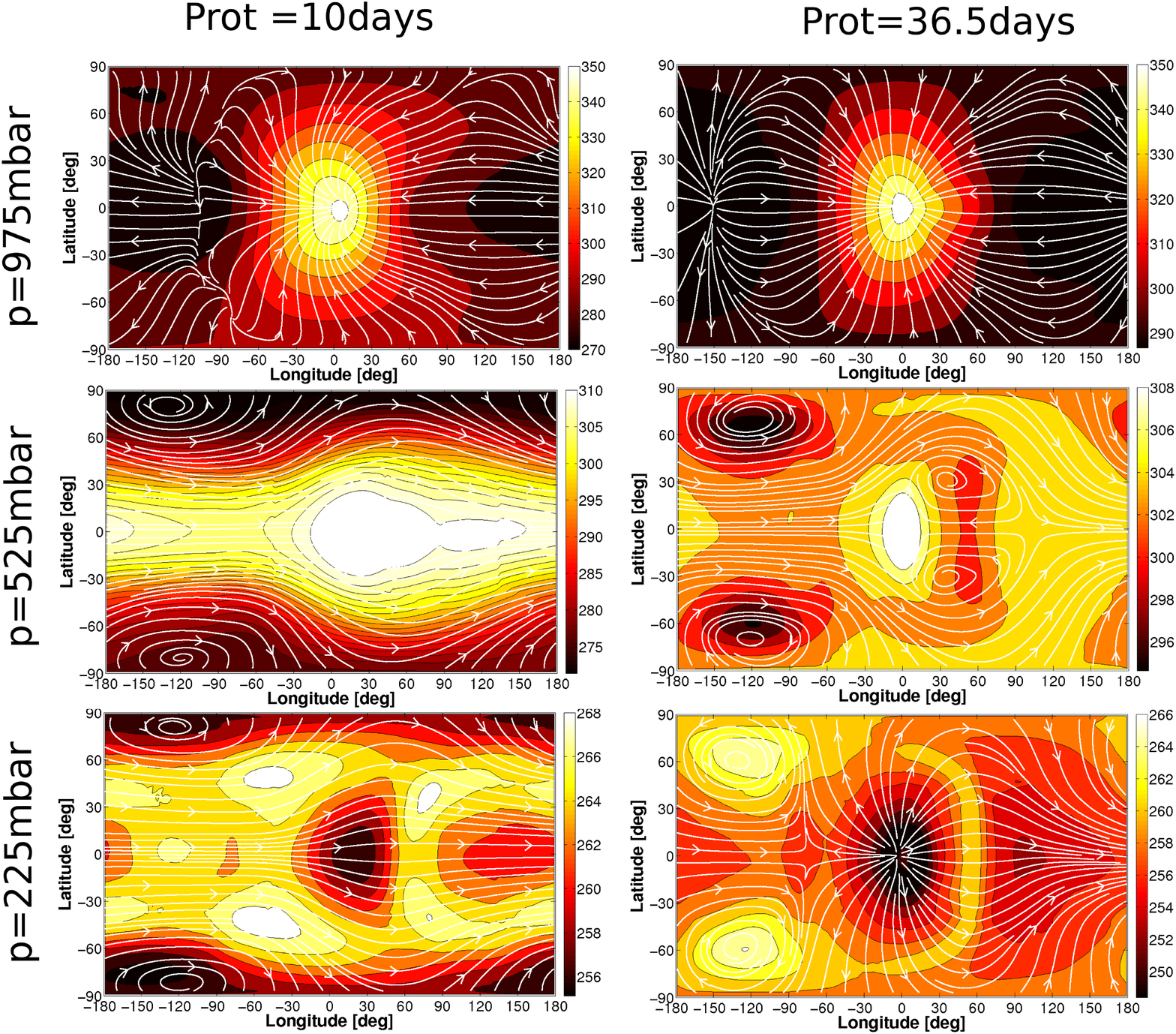}
\caption{Temperatures and streamlines of the flow, averaged over 1000~days, for pressure level $p=975$~mbar (contour interval 10~K), $p=525$~mbar and $p=225$~mbar (contour interval 2~K), from top to bottom. The left panel shows simulations with $P_{rot}=10$~days, the right panel with $P_{rot}=36.5$~days.}
\label{fig:Prot36d_stream}
\end{figure*}

Figure~\ref{fig:Prot36d_stream} shows the temperatures and horizontal flow for the lower, middle troposphere and upper troposphere for $P_{rot}=10$~days and $P_{rot}=36.5$~days. In both cases, the surface flow is directed towards the substellar point at zero longitude and latitude. This flow pattern is consistent with the understanding that the substellar point is dynamically the Earth's tropic equivalent with upwelling and thus converging horizontal flow at the surface and divergence at the top of the atmosphere. The latter is indeed directly visible due to the divergence in horizontal flow for our slow rotating case at $p$=225~mbar embedded in a generally westerly flow that encompasses the substellar point.

For the fast rotating case (left panels in Figure~\ref{fig:Prot36d_stream}), the equatorial superrotation supersedes divergence. The local temperature minimum at $p$=225~mbar at the substellar point, however, indicates the top of the ascending circulation branch. This connection between top horizontal flow and circulation cells is further confirmed by the meridional overturning streamfunction $\psi$ that is defined in the framework of the Eulerian mean, where a flow parameter $A$ is decomposed into a zonal average $\bar{A}$ and a longitudinally varying or eddy part $A'$, thus:
\begin{equation}
A=\bar{A}+A'.\label{ref:eddy_def}
\end{equation}
$\psi$ can then be calculated in pressure coordinates by (see, e.g.,\cite{Holton})
\begin{equation}
\psi(p,\nu) = \frac{2\pi R_P \cos\nu}{g}\int^p_0 \bar{v} \mathrm{d}p
\end{equation}
and has units of kg/s.

\begin{figure*}
\includegraphics[width=0.495\textwidth]{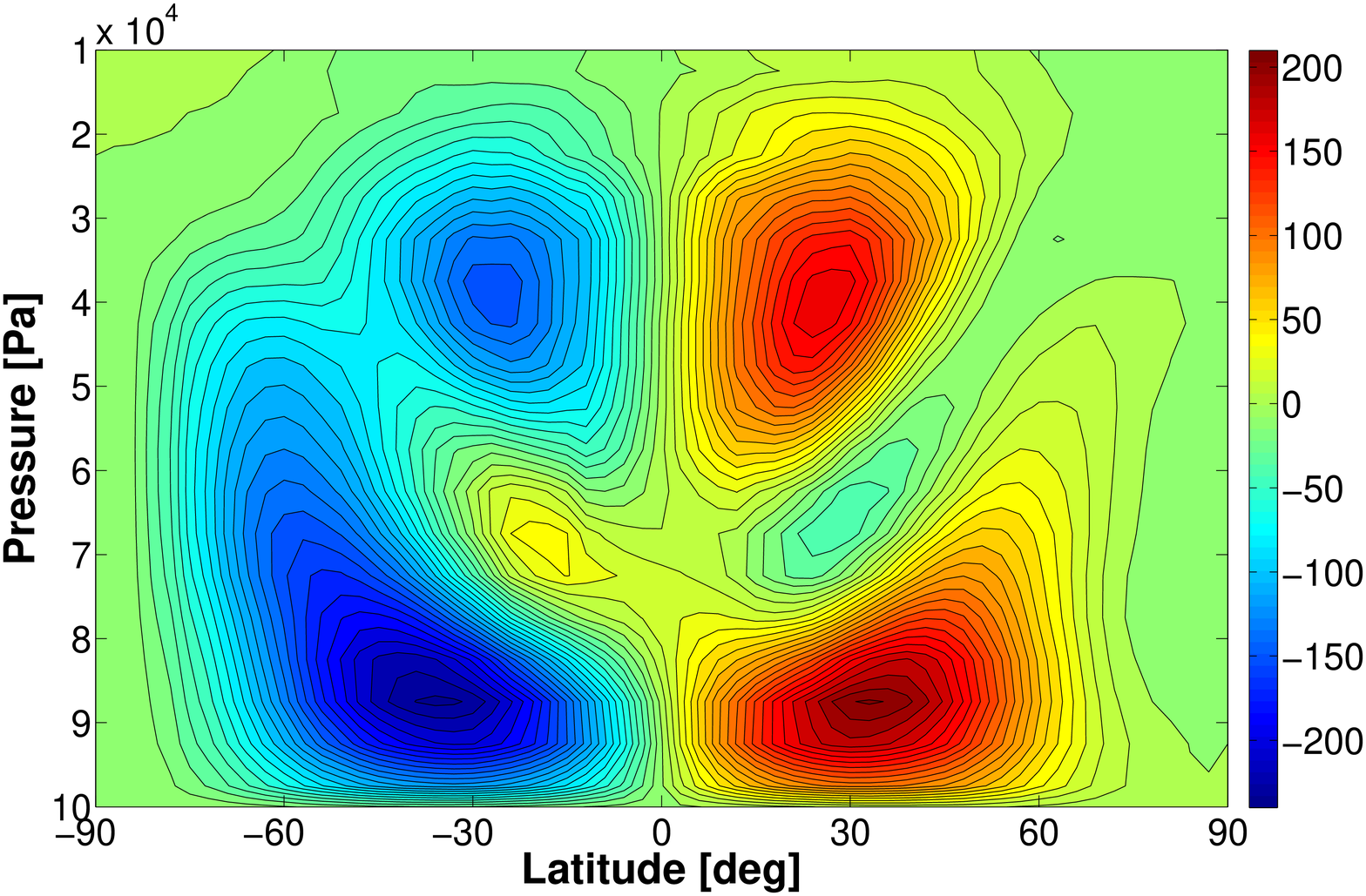}
\includegraphics[width=0.495\textwidth]{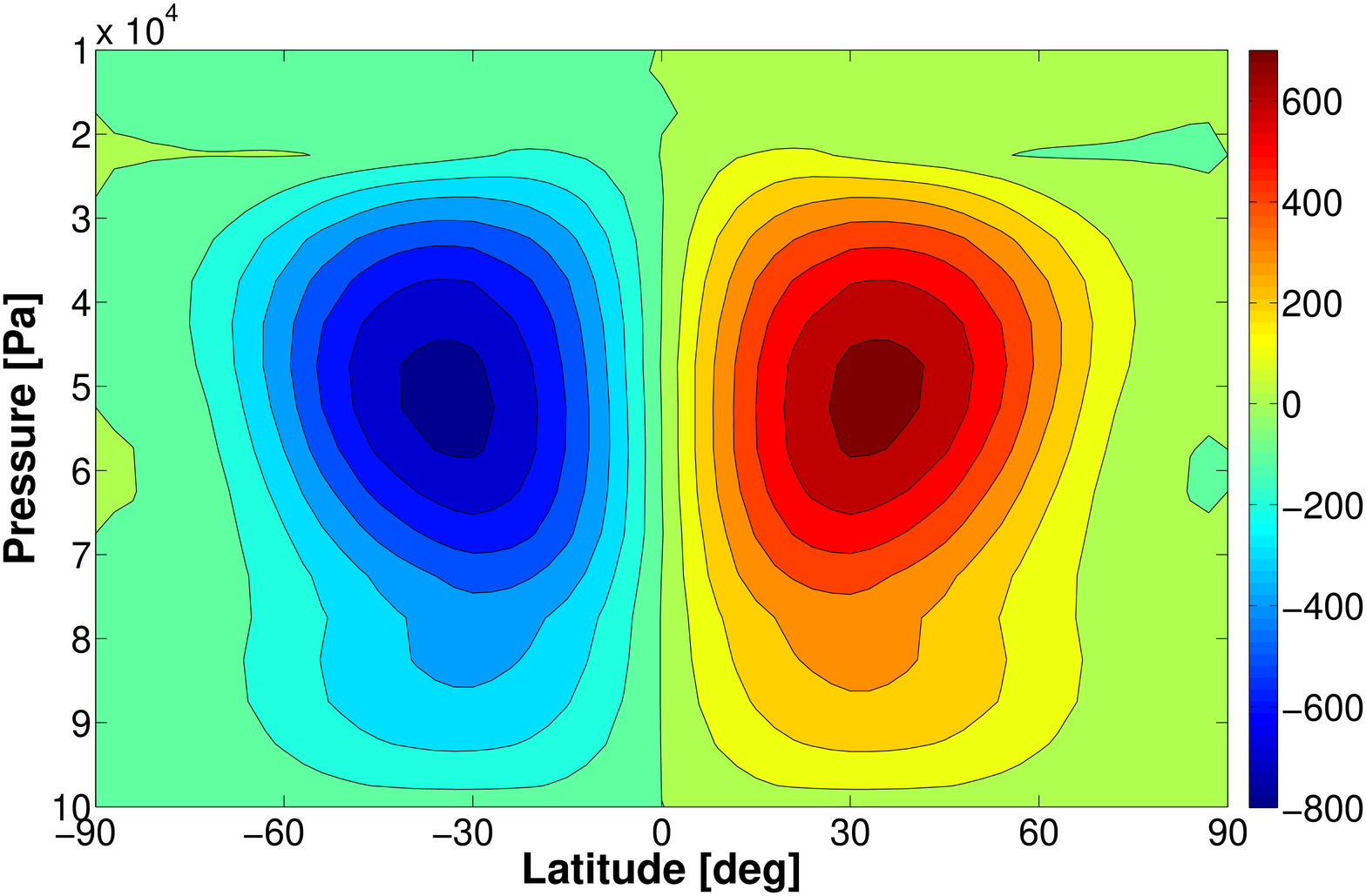}
\caption{Meridional overturning cell in units of $10^9$~kg/s for the fast ($P_{rot}=10$~days, left panel, contour level: $10\times 10^9$~kg/s) and slow rotating case ($P_{rot}=36.5$~days, right panel, contour level: $100\times 10^9$~kg/s). Note that positive values indicate clockwise and negative values counterclockwise circulation.}
\label{fig:Overturn}
\end{figure*}

Figure~\ref{fig:Overturn} plots this streamfunction $\psi$ and shows two large circulation cells, one for each hemisphere, with upwelling at the equator and downwelling at the poles. In the fast rotating case, we see embedded in the large circulation another pair of circulation cells that we will further elaborate upon in the next section.

Our results for the horizontal flow are in agreement with \cite{Merlis2010} who found that slow rotating planets show divergence at the top of the atmosphere and that fast rotation periods lead predominantly to superrotation.

Due to conservation of angular momentum and because we have imposed surface friction, the superrotation in the higher atmosphere is balanced by counterrotation, that is, easterly surface flow. This counterrotation is visible as low-lying negative zonal wind values in the bottom layers of Figure~\ref{fig:Zonal wind}. The zonally averaged zonal wind speed shown there also reveals that the superrotation is not only more prominent in the fast rotating case, it is also stronger than in the slower rotating model. This decrease in superrotation strength with increase in rotation period is in line with the findings by \cite{Edson2011} that the equatorial wind speed continuously drops after transition from the fast regime for $P_{rot}\geq 3$ days. A survey with different rotation periods will show if the anti-cyclonic vortices that appear in the $P_{rot}=36.5$~days model east of the substellar point (right panel of Figure~\ref{fig:Prot36d_stream} for $p$=525~mbar) are a by-product of reduced equatorial zonal wind speed or are actively decelerating the jet.

Furthermore, it is noticeable that the amplitude of the planetary large scale wave is larger for $P_{rot}=36.5$~days compared with the $P_{rot}=10$~days case (Figure~\ref{fig:Prot36d_stream} in particular $p$=525~mbar). This amplitude increase can be understood in terms of length scale of the equatorial Rossby wave of deformation (e.g., \cite{Showman2011}):
\begin{equation}
L_R=\left(\frac{\sqrt{gH}}{\beta}\right)^{1/2},
\end{equation}
where $g$ is the surface gravity, $H$ is the scale height, $\beta=2 \Omega/R_P$ is the meridional variation of the Coriolis parameter $f$ at the equator ($\cos \nu =1$), where $\Omega=2\pi/P_{rot}$ is the planet's rotation rate and $R_P$ the radius, respectively. Obviously, $L_R$ increases with increasing rotation period. Another useful property, in particular, when comparing planets of different sizes is the wavenumber $k_{ER}$ of an equatorial Rossby wave (e.g., \cite{Showman2011}):

\begin{equation}
k_{ER}=\left(\frac{\sqrt{g H}}{2\Omega R_P}\right)^{1/2}.
\end{equation}
This dimensionless number expresses how much of the Rossby wave 'fits' on a planet, where we can assume that the whole planet is filled if $k_{ER}=0.5$ or $L_R=0.5 R_P$. Indeed, \cite{Edson2011} found that the transition from the fast to slow rotating regime -- that is, their $P_{rot}=3-4$~days limit -- coincides with the $k_{ER}= 0.5$-limit, confirming the role of the equatorial Rossby wave as an important driving mechanism for superrotation. For slower rotation, the Rossby wave no longer 'fits' completely on the planet. This connection between Rossby wavenumber and superrotation also explains why \cite{Edson2011} reported a different transition rotation period for dry and wet planets: The temperatures and thus the scale heights are different. However, \cite{Edson2011} also noted that a planet's atmosphere can assume multiple equilibrium states around the transition regime. Table~\ref{tab: Charact_atmo} lists the equatorial Rossby wave number for our two cases ($k_{ER}=1.5$ for $P_{rot}=10$~days and $k_{ER}=2.8$ for $P_{rot}=36.5$~days) and also some other useful characteristics like the maximum zonal wind speed $u_{max}$, the maximum strength of the meridional overturning stream function $\psi_{max}$ and the buoyancy frequency $N=\sqrt{-\rho g^2/\theta \times d\theta/dp}$. The table shows that both cases are in the $k_{ER}>>0.5$ regime. Nevertheless, the dynamics of the mid- and upper troposphere differs quite dramatically even though both show generally superrotating flow. \cite{Showman2011}, on the other hand, assume $k_{ER}=0.5-2$ as a valid range for a Hot-Jupiter-like dynamics regime dominated by an equatorial superrotating jet, which is more in line with our findings that our faster rotating planet is super-rotating whereas our $P_{rot}=36.5$~days-planet with $k_{ER}>2$ exhibits a more divergent flow.

Indeed, the eddy geopotential height $\Phi'/g$ and eddy wind $\vec{v}'$ (calculated via equation~(\ref{ref:eddy_def})) at $p$=200~mbar (Figure~\ref{fig:Eddy}, compared with left panels of Figures~5 and 7 in \cite{Edson2011}) reveal Rossby wave gyres at mid-latitudes and standing waves at the equator similar to the $P_{rot}=120$~h case of \cite{Edson2011} and also comparable to the height fields calculated from the shallow water model by \cite{Showman2011} for the $\tau_{rad}=100$ case. \cite{Showman2011} showed that standing Rossby-Kelvin waves are the underlying mechanism for generating the equatorial superrotation. The slow rotating $P_{rot}=36.5$~days case, on the other hand, shows a mixture between divergent dynamics at the substellar point and standing Rossby and Kelvin waves at other locations that are reduced in strength compared to the faster rotating case. 

Thus, it appears that the $k_{ER}=0.5$ transition denotes the onset of divergence on the top of the atmosphere countering superrotation that decreases in strength until the flow becomes fully divergent for $P_{rot}=100$~days according to \cite{Edson2011} and \cite{Merlis2010}. 

Shallow water models further suggest that not only changes in the rotation period and thus in the Coriolis force, but also in the thermal forcing can lead to a transition between the divergent and superrotation regime for Hot Jupiters \citep{Showman2013} and terrestrial planets \citep{Showmanbook2013}\footnote{Here, the two regimes are called eddy and jet dominated, respectively.}. The results of \cite{Showmanbook2013} (their Figure~17 (a) and (c)) are indeed qualitatively very similar to the horizontal flow of the upper troposphere of our full 3D-model, where strong forcing and thus a small radiative time scale ($\tau_{rad}=0.016$~days) correspond dynamically to our slow rotation simulation model with more divergent flow and weaker forcing ($\tau_{rad}=1.6$~days) to our $P_{rot}=10$~days-simulation with superrotating flow.

The model of \cite{Zalucha2013} has a higher thermal forcing than our model, which suggests a tendency to a more divergent flow. The upper atmosphere levels show instead strong equatorial superrotation. Apparently, the divergence tendency of the strong forcing is overcome by the strong Coriolis force (because of the large planetary radius and fast rotation) that favours superrotating flow. Indeed, \cite{Zalucha2013} found $k_{ER}=0.38$, which places their model firmly in the superrotating regime. The authors note, furthermore, that the equatorial jet is substantially broader than the equatorial Rossby deformation radius, hinting at a deviation from the shallow water model by \cite{Showman2011}.

\subsection{Cyclonic vortices}
\label{sec:vortices}

A further interesting feature is the behaviour of the cyclonic vortices in the mid and upper troposphere. The slow rotating case shows them at mid-latidude ($60^{\circ}$) between the antistellar point and dawn terminator (Figure~\ref{fig:Prot36d_stream}). Furthermore, they are associated with temperature anomalies: They are cold centres in the mid-troposphere and warm centres at the upper troposphere for the slow rotating case. \cite{Edson2011} also report mid-latitude vortices between antistellar point and dawn terminator with temperature anomalies and attribute them to the advection of cooler air toward the substellar point along the equator. This might explain why the centres of the vortices are located more polewards in our $P_{rot}=10$~days simulation compared to the $P_{rot}=36.5$~days simulation: The stronger superrotating jet at the equator pushes them towards the poles.

\begin{figure*}
\includegraphics[width=0.495\textwidth]{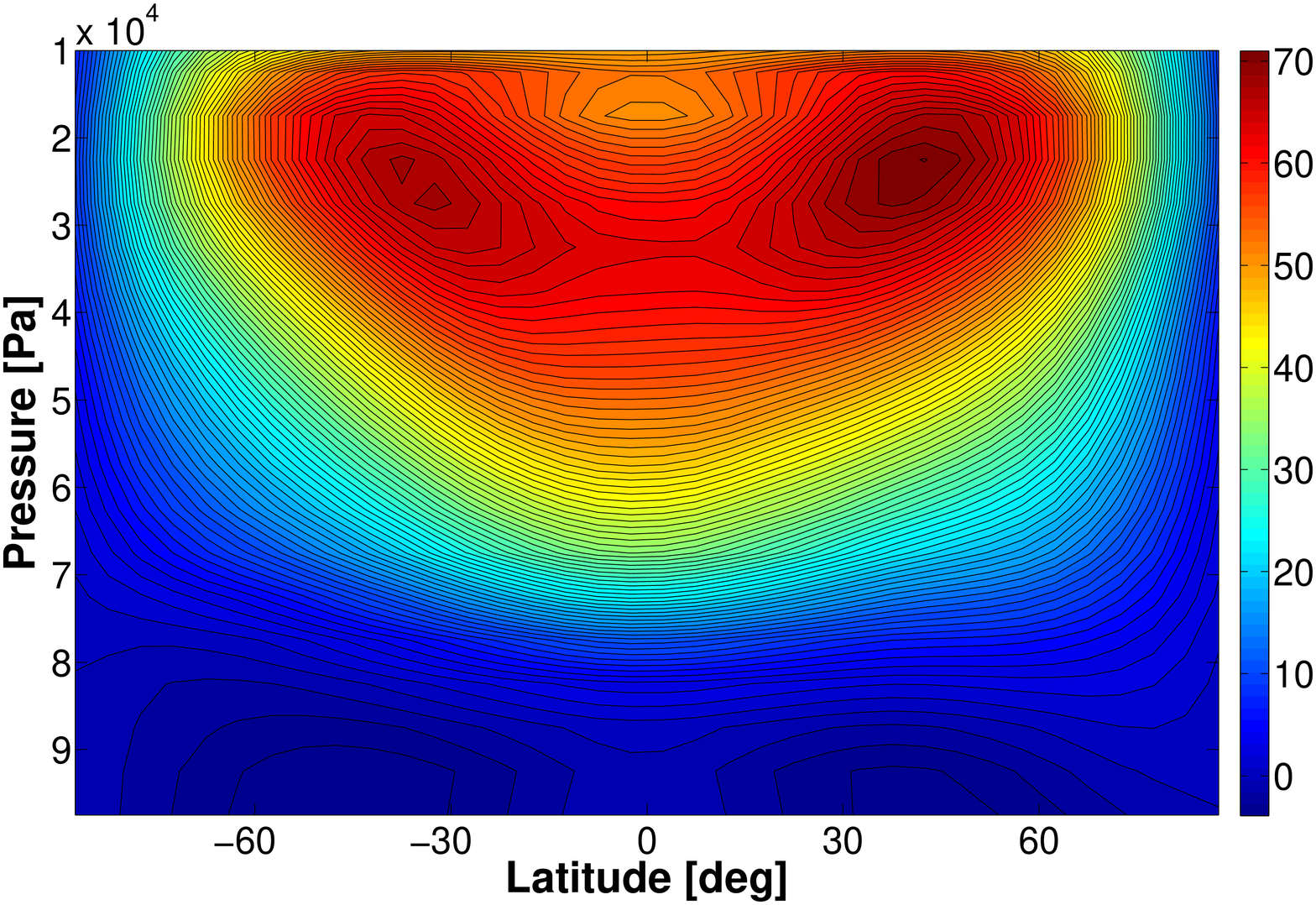}
\includegraphics[width=0.495\textwidth]{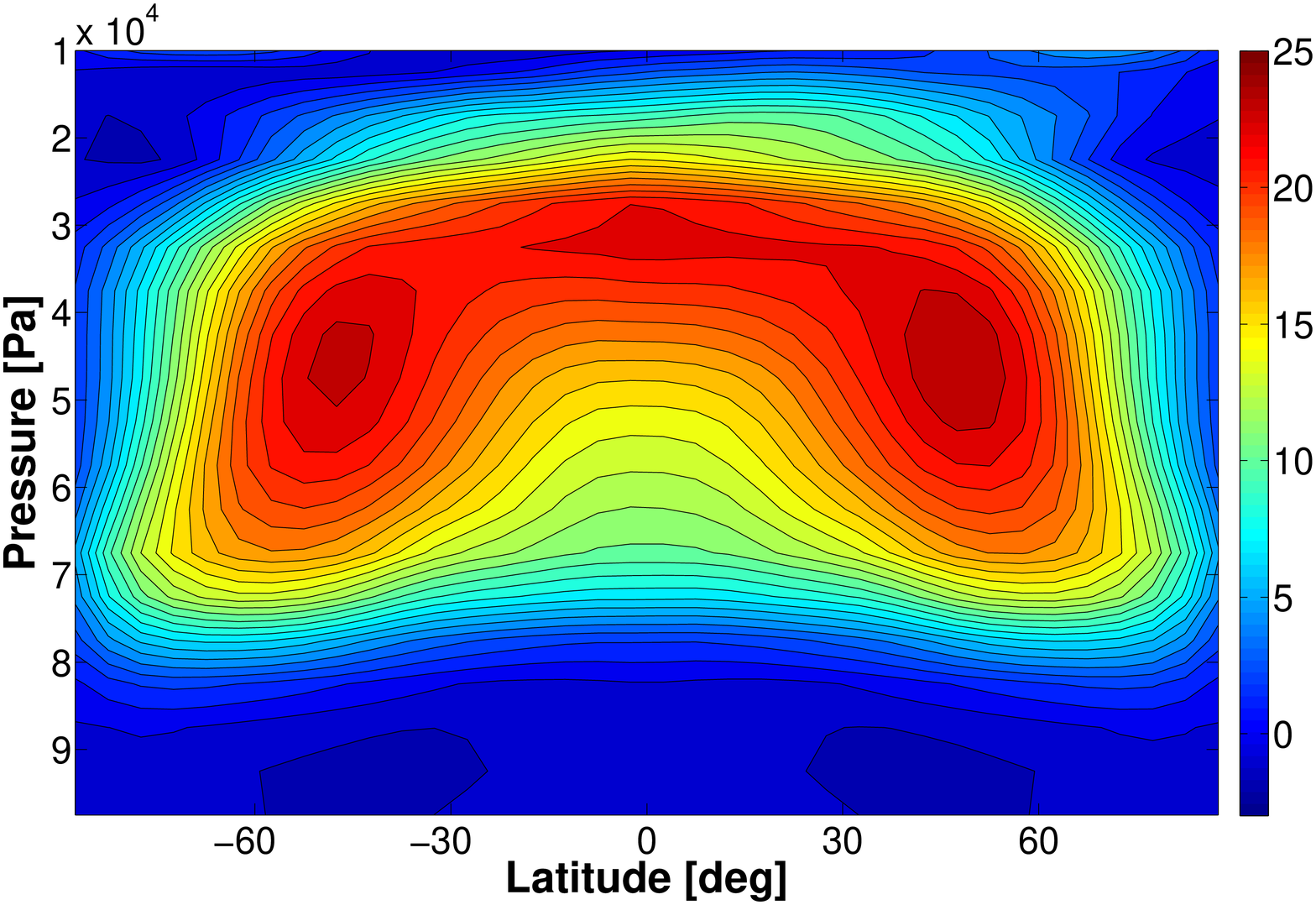}
\caption{Zonal mean of the zonal wind in m/s for the fast ($P_{rot}=10$~days, left panel), and slow rotating case ($P_{rot}=36.5$~days, right panel). Contour levels are 1~m/s.}
\label{fig:Zonal wind}
\end{figure*}

\begin{figure*}
\includegraphics[width=0.495\textwidth]{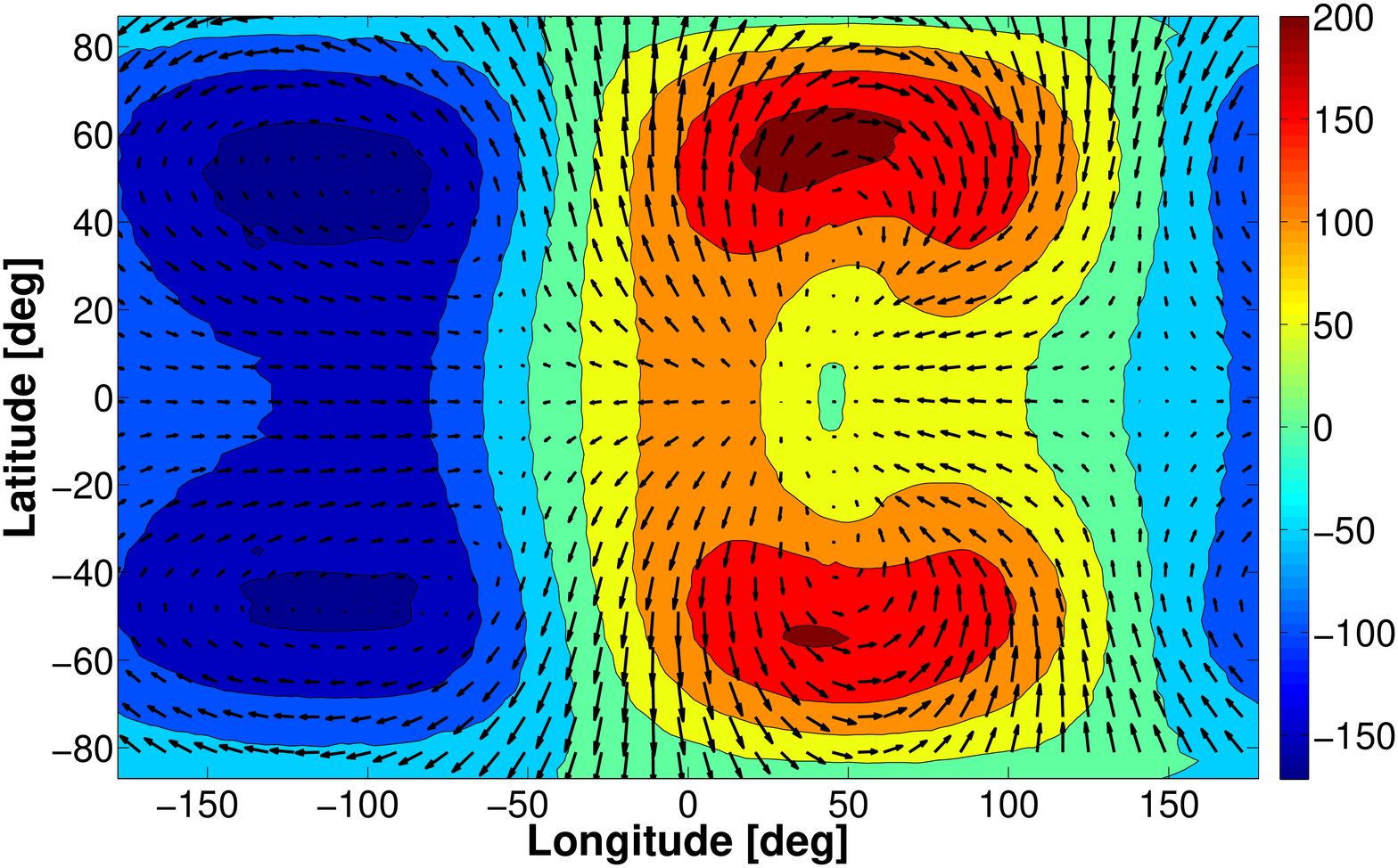}
\includegraphics[width=0.495\textwidth]{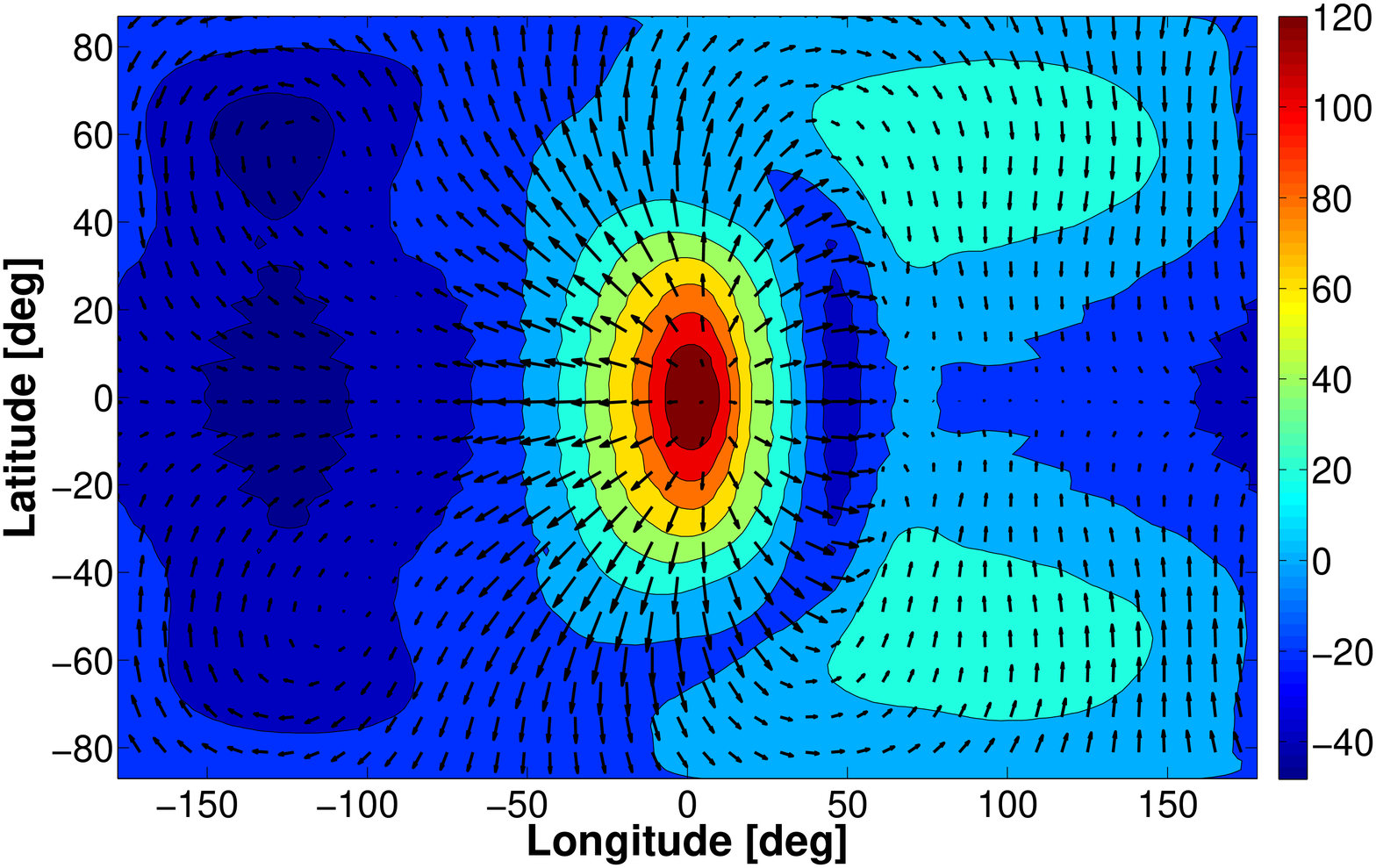}
\caption{Eddy geopotential height in units of m and eddy wind in m/s at $p=225$~mbar for the fast ($P_{rot}=10$~days, left panel), and slow rotating case ($P_{rot}=36.5$~days, right panel). Contour levels are 100~m. The longest wind vectors are 36.57 m/s and 43.54 m/s, respectively.}
\label{fig:Eddy}
\end{figure*}

In contrast, \cite{Edson2011} find the mid-latitude vortices for $P_{rot}=120$~h, which is twice as fast as our fast rotating case $P_{rot}=10$~days, where the vortices are at the poles instead. We note, however, that the  equatorial superrotation in the 120~h model of \cite{Edson2011} is two times smaller compared with our $P_{rot}=10$~days case. Furthermore, we notice that the equatorial edges of the vortices coincide with additional spin-up of the zonal jet in our models (Figure~\ref{fig:Zonal wind}) that show two zonal wind maxima in the upper troposphere (between $p=300$~mbar and $p=200$~mbar) at latitudes $40^\circ$ for $P_{rot}=10$~days and two maxima in the mid-troposphere (between $p=500$~mbar and $p=400$~mbar) at latitudes $50^\circ$ for $P_{rot}=10$~days. Thus we speculate that the vortices, the equatorial jet and the Rossby waves are coupled with each other in our simulations.

The higher location of the maximum zonal winds in our fast rotating case might indicate that the vortices are not only pushed polewards, when compared to the slow rotating case, but also upwards. Furthermore, we see in the slow rotating case at the top of the troposphere at $p=200$~mbar (Figure~\ref{fig:Zonal wind}) the clear emergence of a single maximum at the equator, indicating that the vortices are decoupled from the zonal flow because the dynamics on top of the atmosphere is becoming divergent.

Indeed, we see in the latter $P_{rot}=36.5$~days-case that the vortices become warm temperature anomalies at the top of the atmosphere, whereas the cyclones remain cold air anomalies in the fast rotating case throughout the troposphere (Figure~\ref{fig:Prot36d_stream}, bottom panels). We speculate that these warm temperature anomalies appear because of adiabatic heating. When air that flows from the substellar to the antistellar point reaches the vortices, it falls 'downward' because of the reduced vertical extent of the colder atmosphere in that region. A similar mechanism is responsible for winter polar warming in the middle atmosphere at Mars (e.g., \cite{Forget1999}). 

In the fast rotating case, the cold regions in the middle troposphere are not confined to the vortices but comprise the polewards regions of the superrotating equatorial jet. Therefore, we speculate that we see here warm temperature anomalies that are instead correlated with strong horizontal temperature gradients at the flanks of the warm equatorial jet in the mid-troposphere, because the air at the top of the troposphere `falls down the slopes' of the superrotating mid-troposphere jet and becomes heated (Figure~\ref{fig:Prot36d_stream}, left panels for $p=525$ and $225$~mbar). We will later show that an adiabatic heating mechanism explains also the upper thermal inversions in the vertical temperature profiles.

In contrast, upwelling of the flow at the substellar point should lead to adiabatic cooling and indeed the substellar point is in both our simulations cooler at the top of the troposphere than the surrounding air (Figure~\ref{fig:Prot36d_stream}, bottom panels for $p$=225~mbar). This does not invalidate, however, our argument that once the air has been lifted up, it can stream away towards the nightside and in the process be adiabatically heated under the right conditions. We argue that both processes may happen at the upper troposphere. Furthermore, it should be noted that the involved temperature differences are relatively small: $\Delta T=12-16~K$.

\cite{Edson2011} also describe cyclones at the same location as in our $P_{rot}=36.5$~days model with the difference that they extend down to the surface. \cite{HengVogt2011} also showed mid-latitude vortices between antistellar point and dawn terminator that can extend down to the surface if the friction and radiative timescales taken from the \textit{HS94} are increased by a factor of $30$ for identical planetary parameters than the ones we used here. Apparently, differences in surface friction prescriptions change the vertical extent of the cyclones. It is interesting to note that \cite{Edson2011} don't see our zonal wind maxima, but this difference in zonal wind strength versus latitude might be attributed to a different circulation regime. They have two meridional circulation cells per hemisphere instead of one. \cite{HengVogt2011}, on the other hand, do exhibit maxima of zonal wind speeds at mid-latitude for their nominal model with $P_{rot}=36.5$~days. These zonal wind maxima are, however, not addressed in their study.

\begin{table*}
\caption{Some atmospheric characteristics of the atmospheres of the modeled planets}
\begin{tabular}{|l|c|c|c|c|c|c|}
$P_{rot}$ & $\beta [m^{-1}s^{-1}]$ & $N$ [$s^{-1}]$& $H$ [km] & $k_{ER}$ & $u_{max} [m/s]$ & $\psi_{max}[10^9 kg/s]$\\
\hline
10 days & $1.57\times 10^{-12}$ & 0.0113 & 6.23  & 1.5 & 70 & 221\\
36.5 days & $4.31\times 10^{-13}$ & 0.0130 & 6.65 & 2.8 & 23 & 643\\\
\end{tabular}
\label{tab: Charact_atmo}
\end{table*}

\subsection{Circulation}
\label{sec:circulation}

In general, both our simulations show two giant circulation cells, one for each hemisphere. The presence of only two circulation cells is in line with the understanding that our atmospheres are in a rather slow rotation regime with $P_{rot}\geq 10$~days. Indeed, most authors that have investigated tidally locked slow rotating planets find two circulation cells: \cite{Joshi1997}, \cite{Joshi2003}, and \cite{Merlis2010}. Interestingly, \cite{Edson2011} report consistently four circulation cells even for $P_{rot}=100$~days which is actually quite puzzling as one would expect already from investigations of planets with non-tidally locked Earth-like forcing the emergence of a single circulation cell per hemisphere that extends up to the poles. E.g., \cite{Navarra2002} found the transition to a single hemisphere circulation cell for $P_{rot}\geq 144$~h.

The difference in the number of circulation cells between our simulations and those of \cite{Edson2011} also explains why their circulation strength is reduced by one order of magnitude when compared to our results. They got $\psi_{max}\propto 10\times 10^9$~kg/s, whereas our circulation strength yields $\psi_{max}\propto 200-600 \times 10^9$~kg/s (Table~\ref{tab: Charact_atmo}). Still, \cite{Joshi1995} have $\psi_{max}\propto 80\times 10^9$~kg/s, which is a factor of 2.5 and 7.5 smaller, respectively. They let, however, the nightside of their $CO_2/H_2O$ atmosphere cool towards the condensation temperature of $CO_2$ which is about 150~K and thus higher than our 77~K nitrogen condensation temperature. Therefore, the difference in circulation strength may be attributable to differences in nightside temperature forcing. To sum up, our simulations agree in many features with the results from the study of \cite{Edson2011}, but differ greatly in something as fundamental as the number of circulation cells. This alone shows that additional investigations and intercomparisons are warranted.

\begin{figure*}
\includegraphics[width=0.495\textwidth]{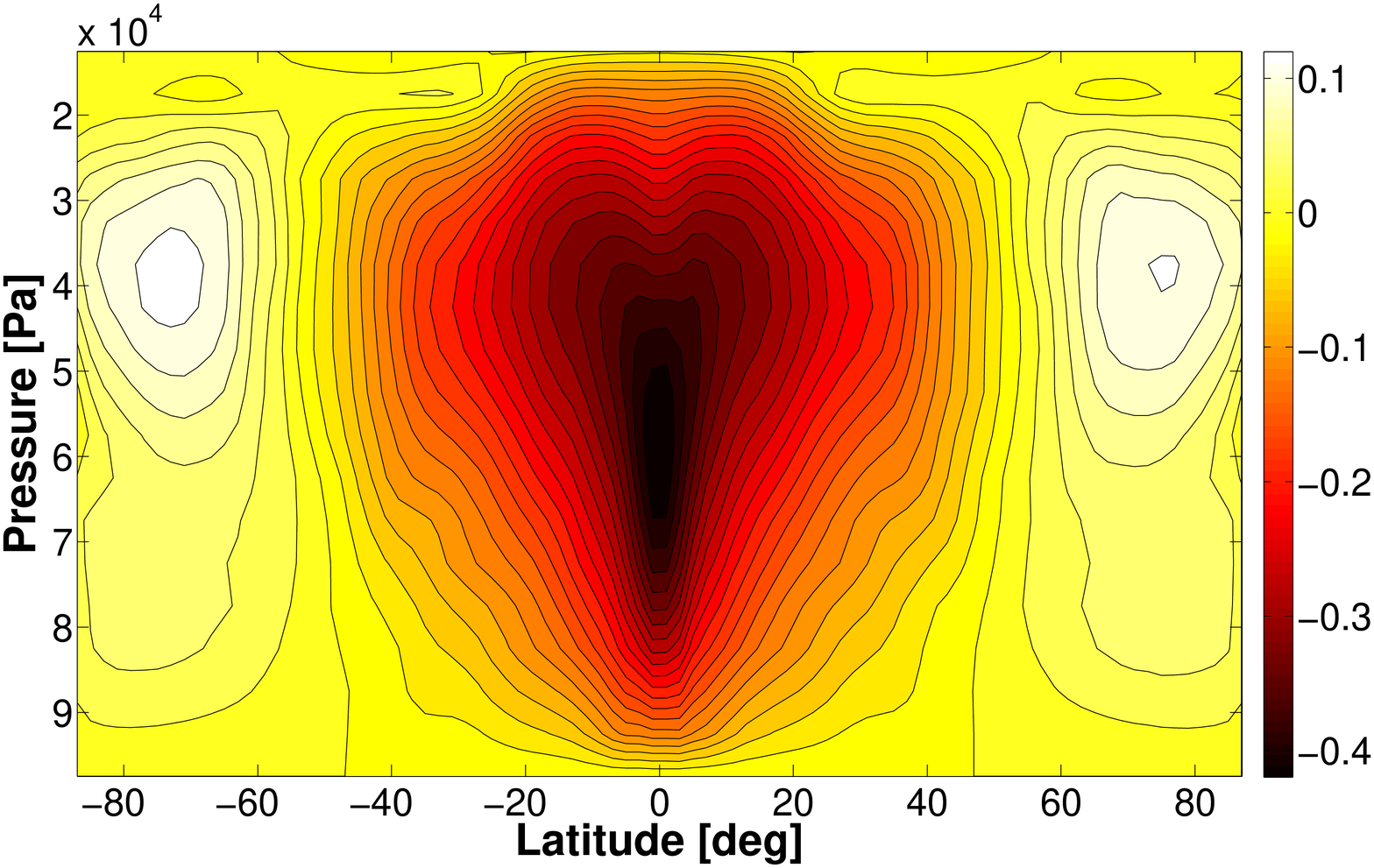}
\includegraphics[width=0.495\textwidth]{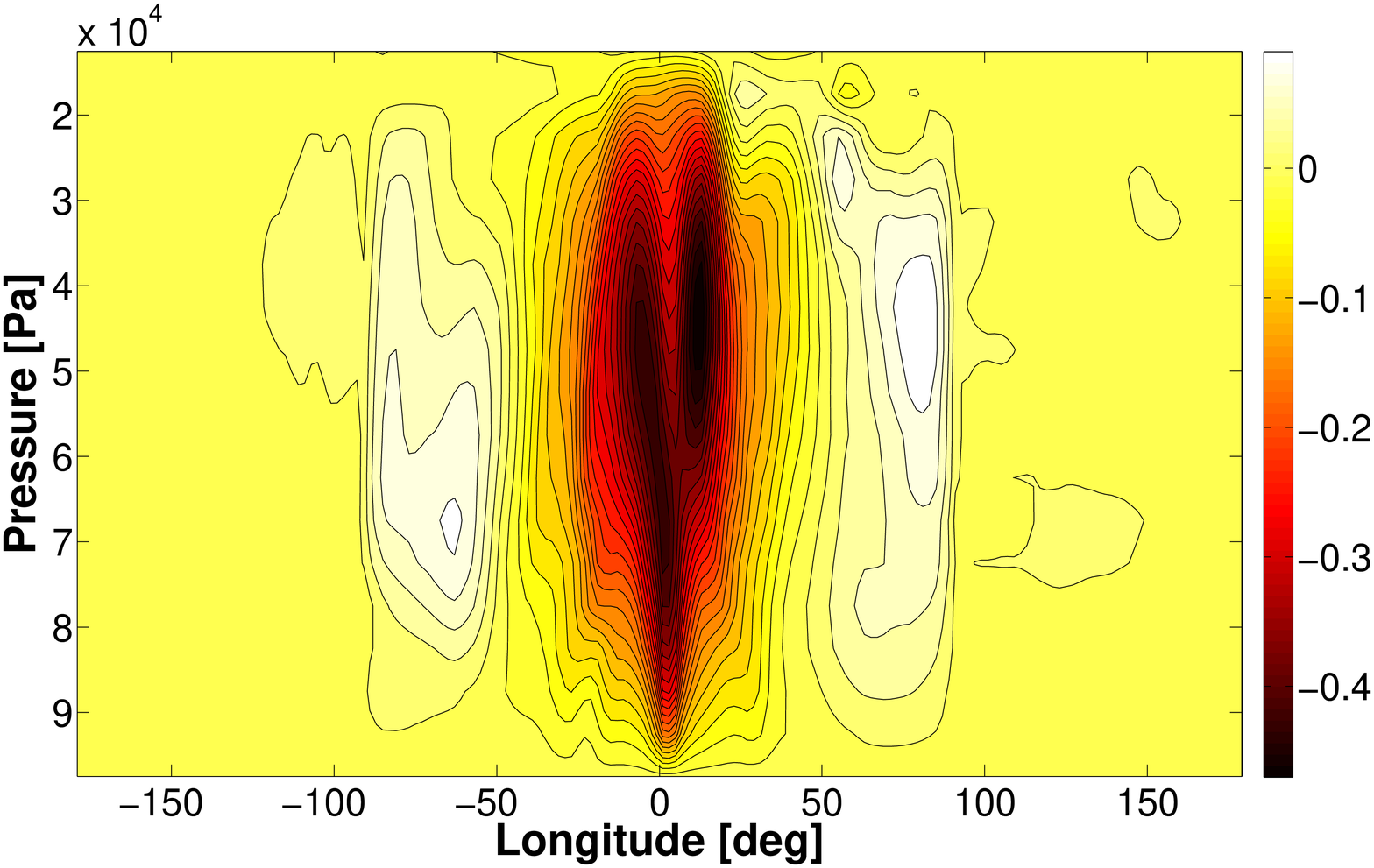}
\includegraphics[width=0.495\textwidth]{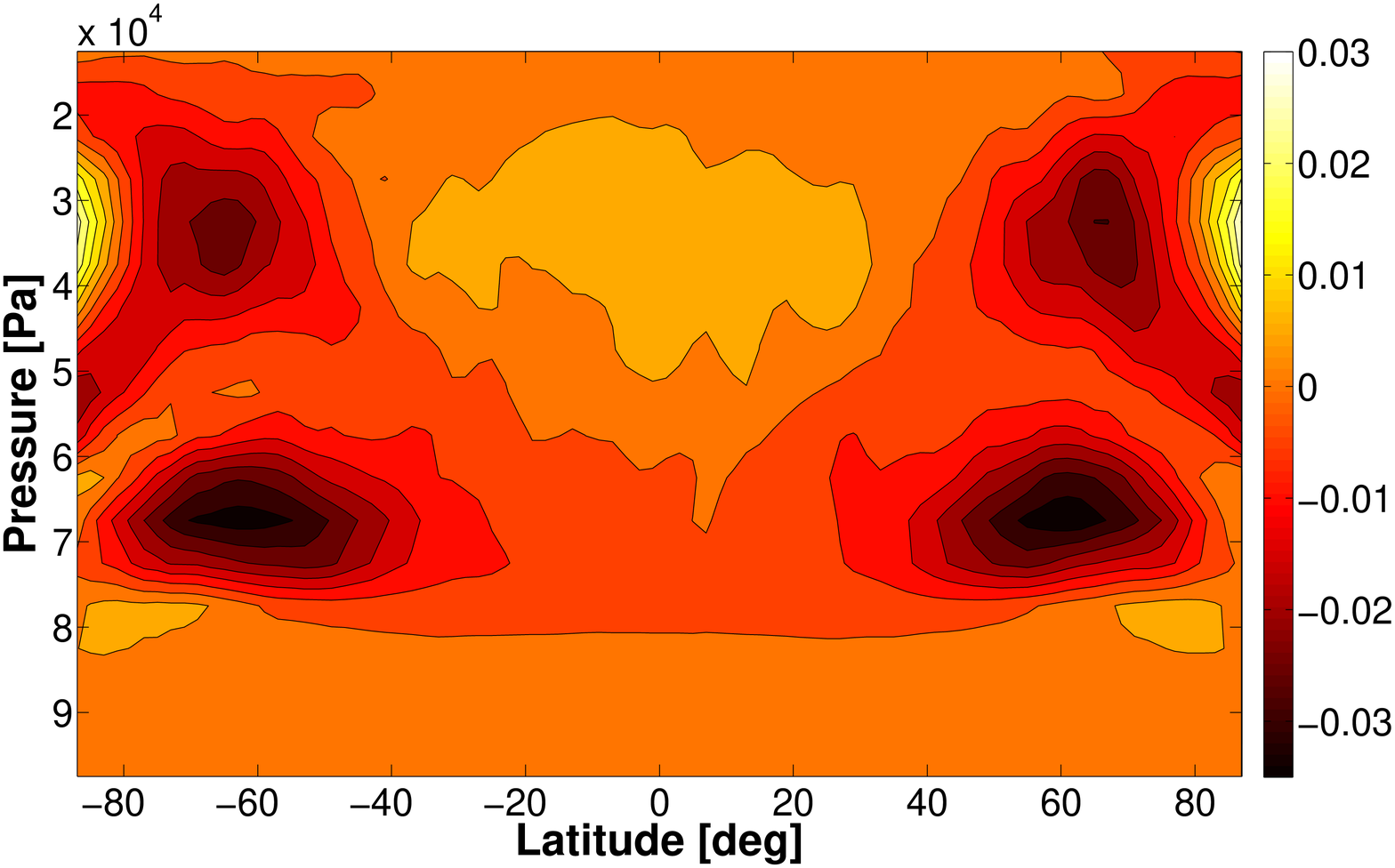}
\includegraphics[width=0.495\textwidth]{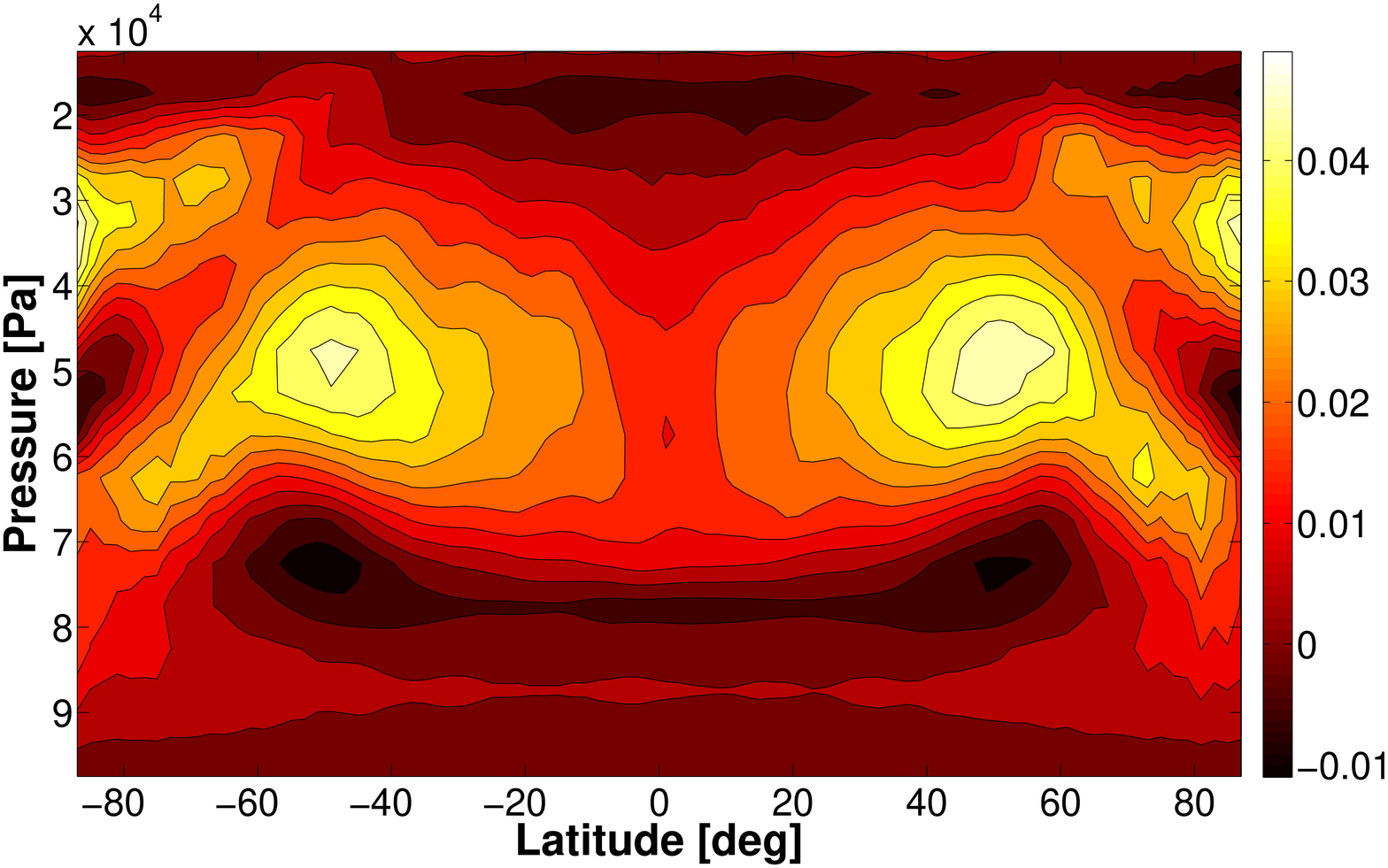}
\caption{Vertical slices of vertical velocity in units of [Pa/s] for $P_{rot}=36.5$~days at substellar point (upper left panel), along the equator (upper right panel), at antistellar point (lower left panel) and at dawn terminator (lower right panel). Negative values denote upward, positive downward motion. Contour levels are 0.02 Pa/s in the upper panels and 0.005 Pa/s in the lower panels.}
\label{fig: W_upwelling_36d}
\end{figure*}

\begin{figure*}
\includegraphics[width=0.495\textwidth]{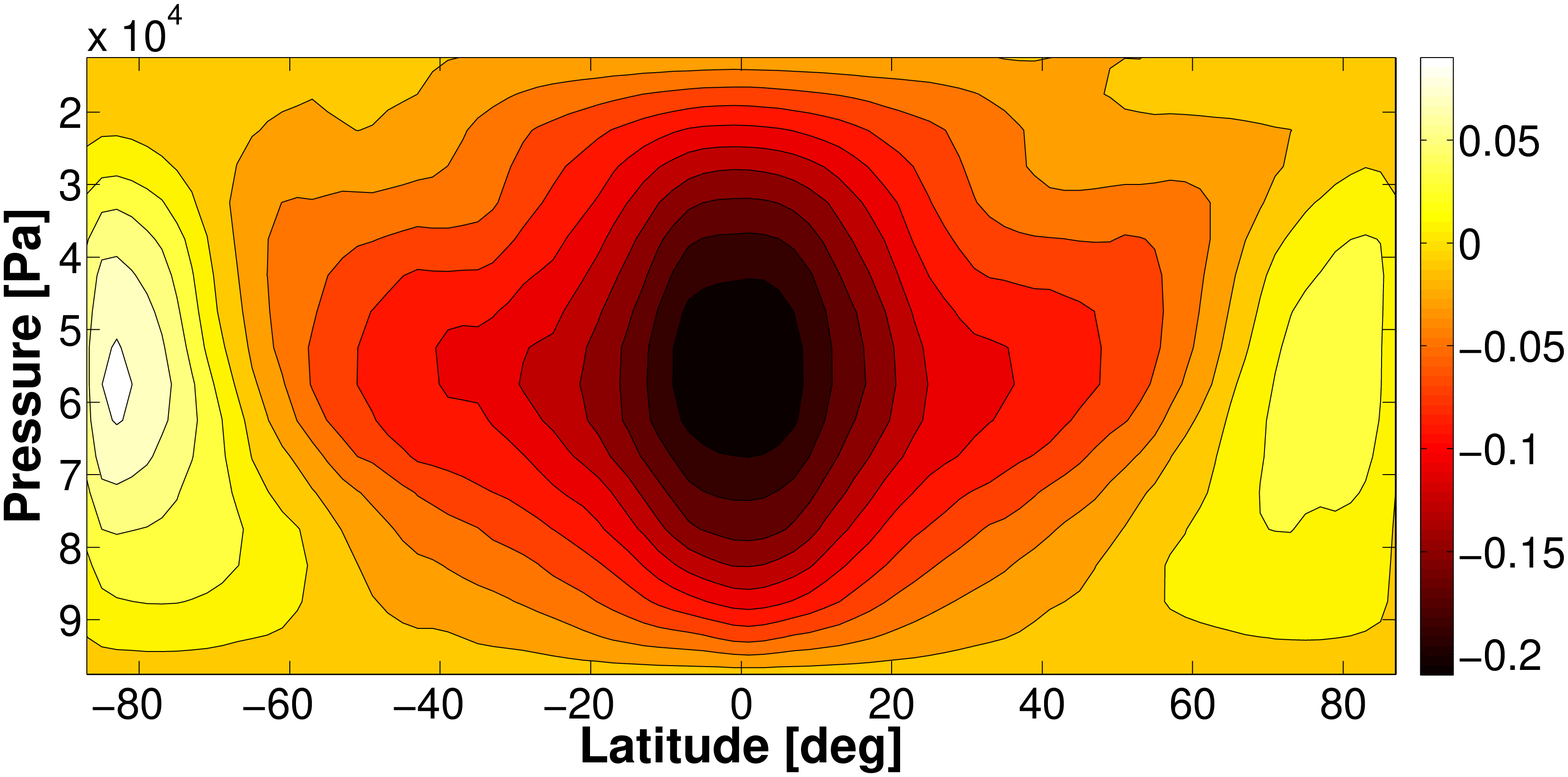}
\includegraphics[width=0.495\textwidth]{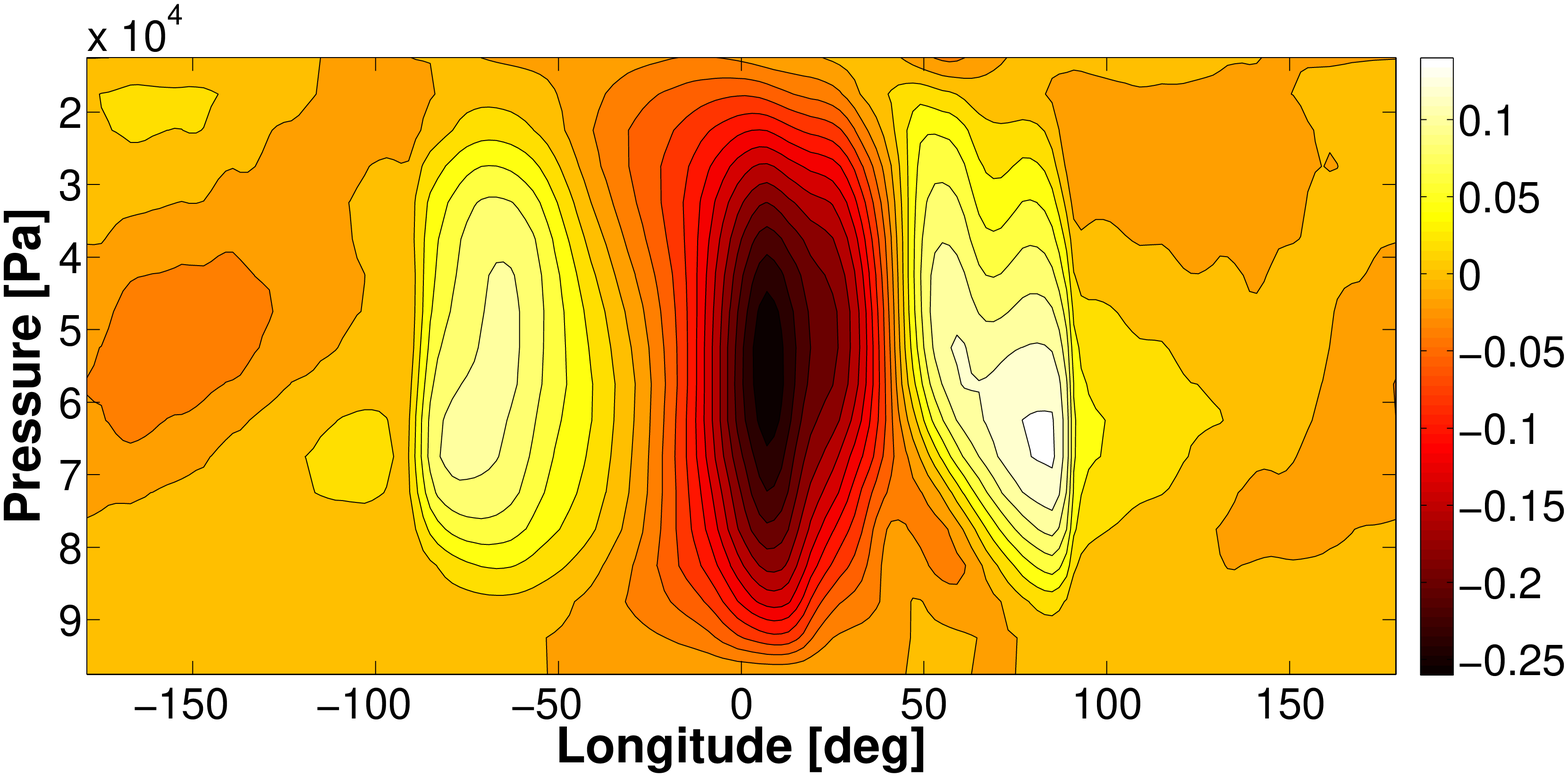}
\includegraphics[width=0.495\textwidth]{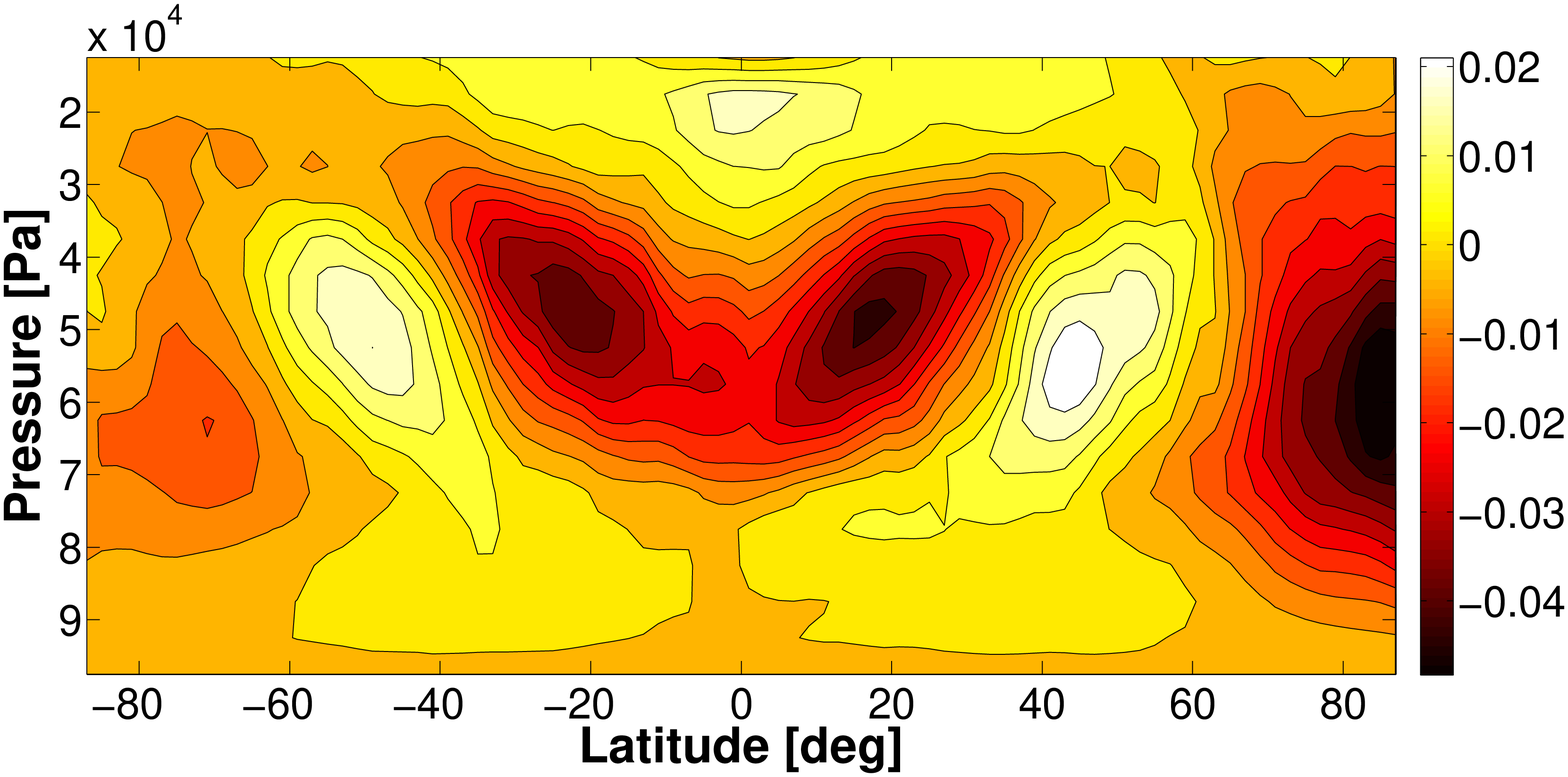}
\includegraphics[width=0.495\textwidth]{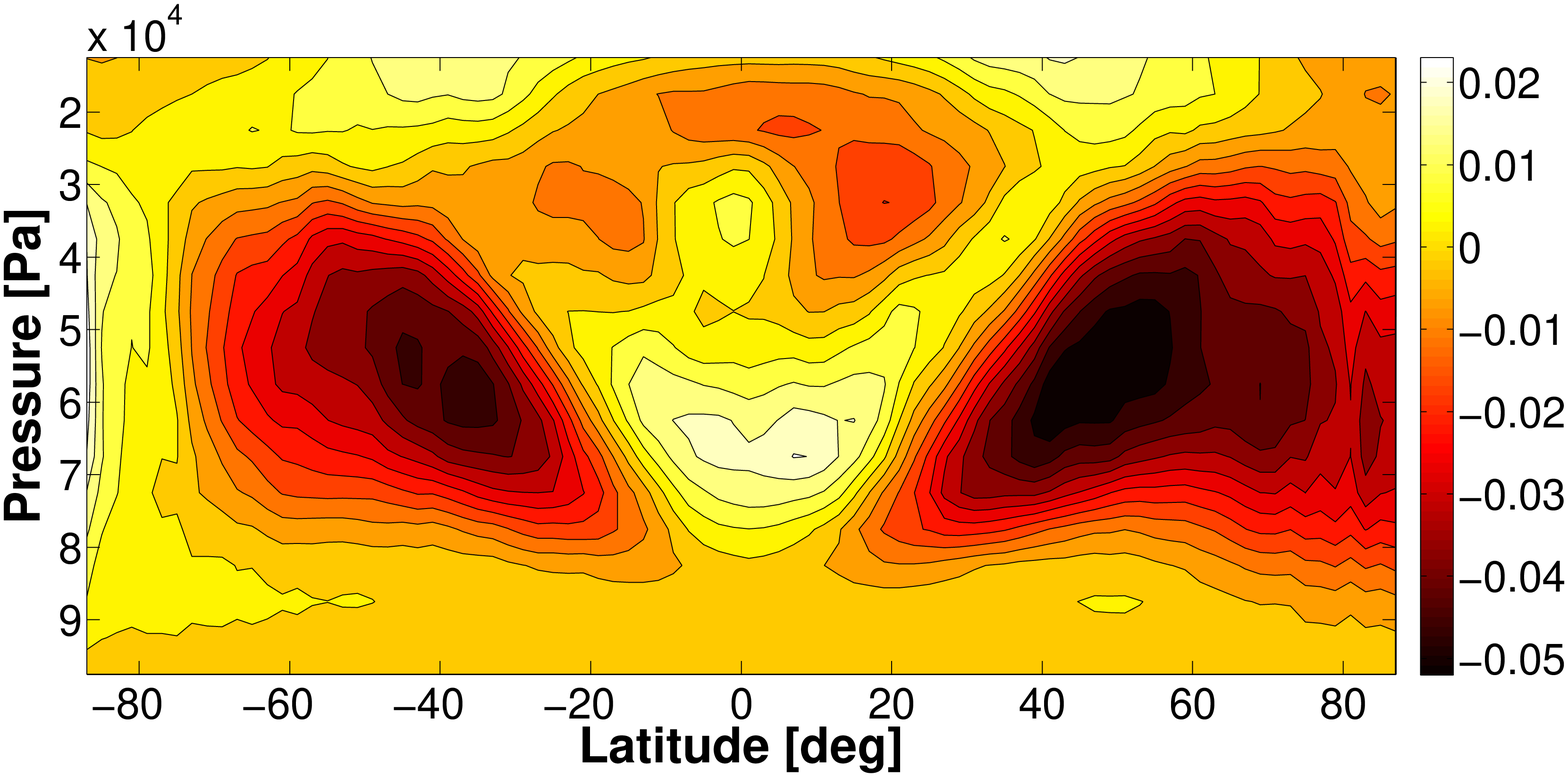}
\caption{Vertical slices of vertical velocity in units of [Pa/s] for $P_{rot}=10$~days at substellar point (upper left panel), along the equator (upper right panel), at antistellar point (lower left panel) and at dawn terminator (lower right panel). Negative values denote upward, positive downward motion. Contour levels are 0.02 Pa/s in the upper panels and 0.005 Pa/s in the lower panels.}
\label{fig: W_upwelling_10d}
\end{figure*}

\begin{figure}
\includegraphics[width=0.45\textwidth]{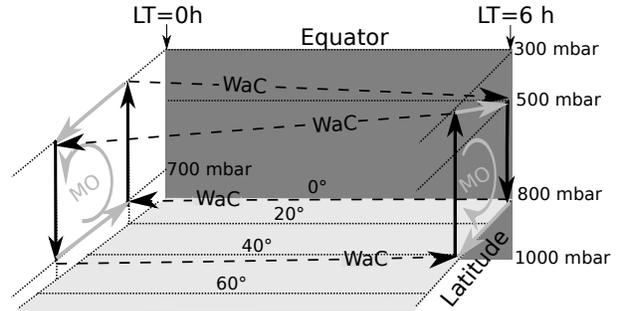}
\caption{Schematic view for the two Walker-like circulations (dashed arrows, WaC) between anti-stellar point (LT= 0 h) and dawn terminator (LT= 6 h) for the $P_{rot}=10$~days simulation in the north hemisphere. The black arrows denote upwelling and downwelling, as seen in Figure~\ref{fig: W_upwelling_10d}. Grey arrows show the resulting  meridional circulation (MO) cells.}
\label{fig: Walker}
\end{figure}

Another interesting feature in the circulation of our modelled Super-Earth atmospheres, for which we can find no counterpart in previous studies for terrestrial exoplanets, is the emergence of a smaller counter rotating meridional cell (directed from high latitudes towards the equator) in mid-troposphere in the $P_{rot}=10$~days case. Vertical cross sections of the vertical winds are very useful in this context and these are shown at substellar, antistellar, the terminators and along the equator for slow rotation in Figure~\ref{fig: W_upwelling_36d} and fast rotation in Figure~\ref{fig: W_upwelling_10d}. They confirm that the overall circulation is driven by flow from the substellar point towards the poles and nightside (Figures~\ref{fig: W_upwelling_36d} and~\ref{fig: W_upwelling_10d}, upper panel). There is strong localized upwelling at the antistellar point and weaker general downwelling elsewhere with centers at the poles and the terminators, in agreement with the results of \cite{Joshi1997}. But there appears to be in addition a Walker-like circulation in the mid-troposphere parallel to the equator with upwelling at high latitudes at the antistellar nightside and corresponding downwelling at the terminator. This Walker circulation is more dominant in the faster rotating $P_{rot}=10$~days case (Figure~\ref{fig: W_upwelling_10d}, lower panels) that shows upwelling at small latitudes at the antistellar nightside and corresponding downwelling at the terminator. Furthermore, a second Walker circulation cell in the other direction is present for the fast rotating case: with upwelling at higher latitudes and corresponding downwelling at the nightside. We speculate that the interplay between these two cells results in a net-meridional circulation counter-rotating to the large cells. The branches of the two Walker circulations at the dawn terminator (LT=6~h) apparently result in a local meridional circulation cell (Figure~\ref{fig: Walker}) with opposite direction to the large hemispheric cell (Figure~\ref{fig:Overturn}, left panel). In the slow rotating case, the connection between mid-troposphere downwelling at dawn terminator and upwelling at the nightside is not entirely clear (Figure~\ref{fig: W_upwelling_36d}, lower panels). In addition, the origin of the surface upwelling at dawn terminator is puzzling (Figure~\ref{fig: W_upwelling_36d}, bottom right panel) and has no counterpart in the fast rotating case (Figure~\ref{fig: W_upwelling_10d}, bottom right panel). 

\subsection{Temperatures and habitability}
\label{sec: habitability}

When it comes to evaluating the possible habitability of a planet, surface temperatures are of the uttermost importance as they determine if and where liquid water is possible. When we inspect the average nightside temperatures in our model, it is striking that they are relatively warm (Table~\ref{tab: Temp} and Figure~\ref{fig:Prot36d_stream}), although we drive the nightside temperature to a nitrogen condensing equilibrium temperature. These moderate nightside temperatures despite tidal locking confirm once again the statement by \cite{Joshi1997} that a sufficiently dense atmosphere ($p\geq 100$~mbar) distributes heat efficiently from the illuminated dayside to the nightside. Indeed, given the high timescales for nightside cooling ($\tau_{rad}$=813~days), it is not surprising that dynamical heating counteracts any strong cooling tendencies.

\begin{table*}
\caption{Temperature of our models}
\begin{tabular}{|l|c|p{3cm}|p{3cm}|}
$P_{rot}$ & $T_{s}$ [K] & Minimum nightside surface temperature [K]& Maximum dayside surface temperature [K] \\
 & & (1000 days average)& (1000 days average)\\
\hline
10 days & 298.78& 272.21& 352.18\\
36.5 days & 292.61& 286.72& 357.62\\
\end{tabular}
\label{tab: Temp}
\end{table*}

Still, our nightside temperatures barely drop below the freezing temperature of water (273.15~K) for $P_{rot}=10$~days and are well above the freezing temperature for $P_{rot}=36.5$~days. Indeed, while our dayside temperatures are comparable to the dry case of \cite{Edson2011}, their nightside temperatures are lower by more than 100~K.

On the other hand, our nightside temperatures are comparable to those in the model of \cite{Joshi1997}, but their dayside temperatures are in the order of 300~K, which is substantially lower (by approx. 50~K) than our temperatures, even though they prescribed the same flux surface distribution than in our model ($F\propto I_0 \cos\phi \cos\nu$) and a maximum surface equilibrium temperature $T_{s,max}=390$~K versus $T_{s,max}=410$~K used in our forcing, which is only 20~K warmer (Figure~\ref{fig:Earth_forcing}, calculated with equations~(\ref{eq: eff_ds}), (\ref{eq: eff_ds_max}), and (\ref{eq: thin})). The Earth-like aquaplanet with solar-like irradiation used by \cite{Joshi2003}, on the other side, yielded minimum nightside temperatures of 200~K and maximum dayside temperatures of 320~K, in general agreement with the wet models investigated by \cite{Edson2011}, which is not surprising since in both cases Earth-climate models were adapted to the tidally locked forcing scenario.

Therefore, at the current state it has to be concluded that there is a general agreement that tidally locked terrestrial planets allow for liquid water for solar-like insolation, in particular, at the substellar point because dynamics efficiently distributes heat from the day to the nightside. The substellar point, in addition, appears to be a dynamical Earth tropic analogue with upwelling and thus possible cloud formation and precipitation, making it an ideal target for future searches for possible life.

However, there seems to be great disagreements in how this heat transport is achieved (with one or two circulation cells for terrestrial planets and $P_{rot}\geq 10$~days) and how strong this transport is. Comparison of different studies yields vast differences for nightside temperatures that can vary between $156$~K (dry model of \cite{Edson2011}-- which would actually allow for $CO_2$ outfreezing -- and $270$ to $290$~K (our model and \cite{Joshi1997}). The main difference in this context appears to be how the thermal forcing at the nightside is treated. Thus it is no surprise that dry models that prescribe condensation temperatures at the nightside, like our model and \cite{Joshi1997}, tend to agree, while others that impose in addition the presence of an ocean, like \cite{Joshi2003} and \cite{Edson2011}, yield general agreement with each other.

\begin{figure*}
\includegraphics[width=0.495\textwidth]{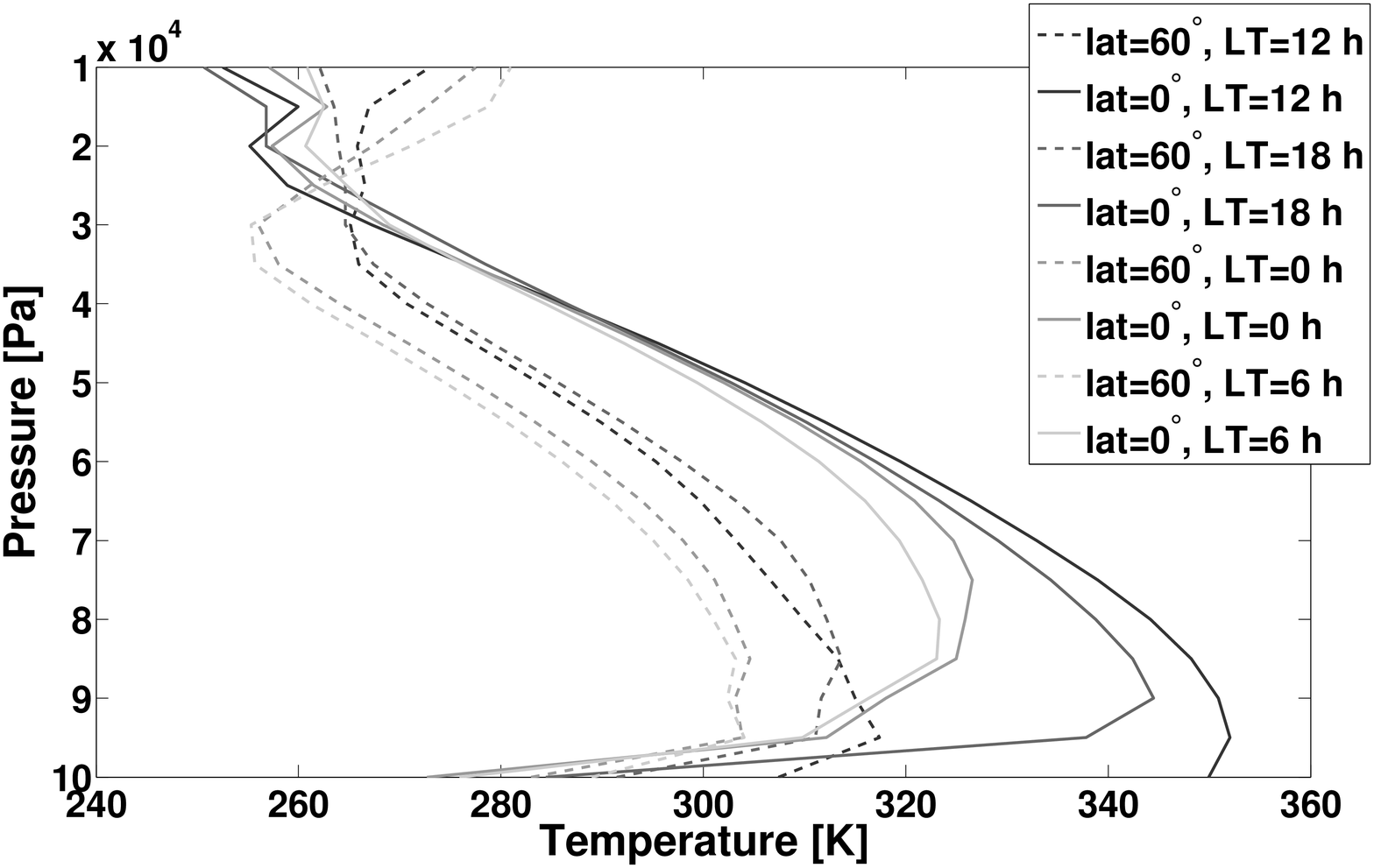}
\includegraphics[width=0.495\textwidth]{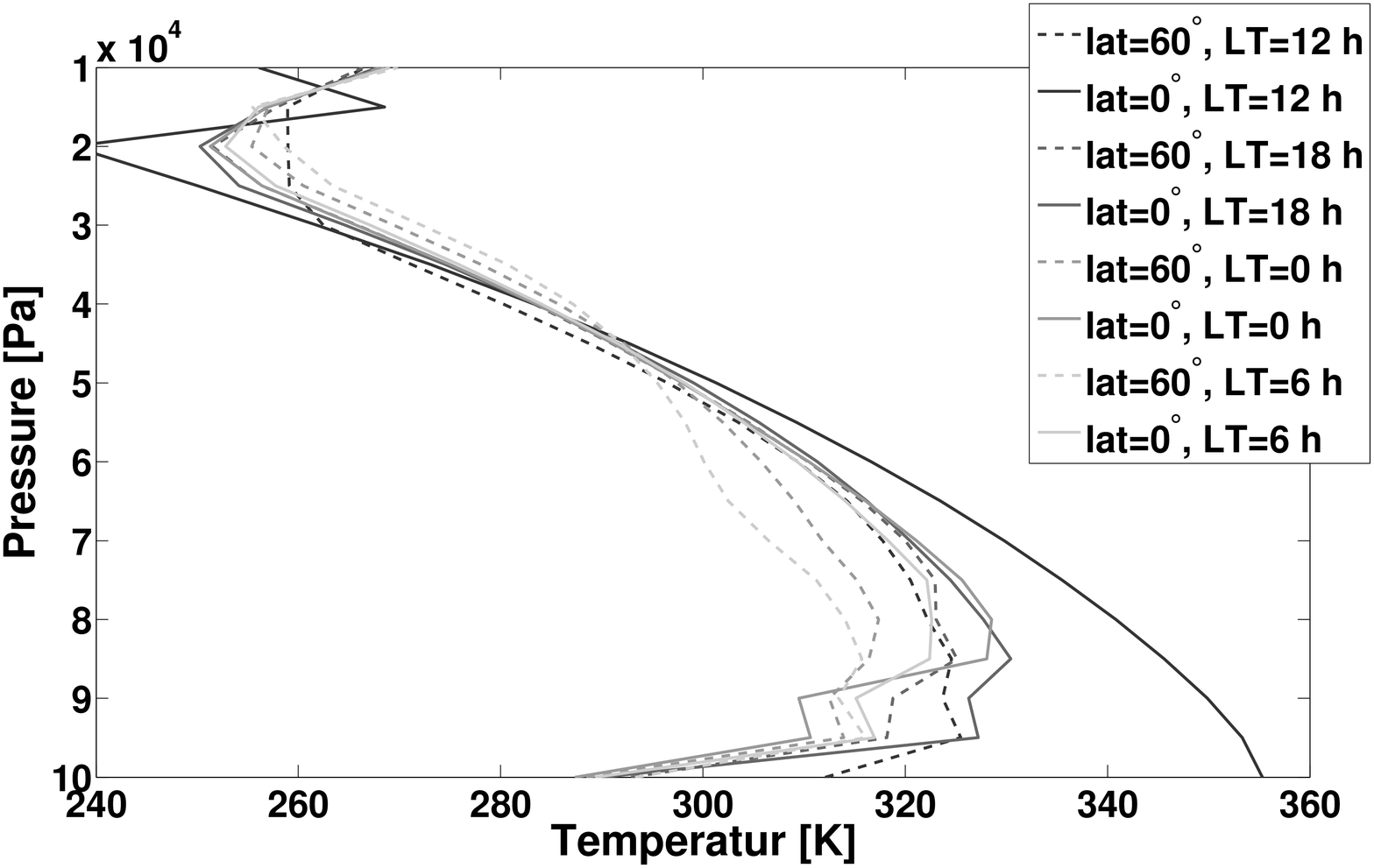}
\caption{Temperature profiles for $P_{rot}=10$~days (left panel) and for $P_{rot}=36.5$~days (right panel). The data are averaged over 1000 days. Temperature profiles at two latitudes, equator (solid lines) and $60^\circ$ (dashed lines), are shown. The longitude positions of the profiles are indicated in different shades of grey (see legend) as local time LT, where LT=0 h is the antistellar nightside and LT=12 h is the substellar dayside. LT=6 h and LT=18 h are the dawn and dusk terminator west and east from the substellar point, respectively. The plots are comparable to those in \citet{Zalucha2013}.}
\label{fig:Temp_profile}
\end{figure*}

While in our model, the nightside temperatures are well above the prescribed nitrogen condensation temperature, there is another feature that suggests that the nightside is indeed cooling down. The vertical temperature profiles in Figure~\ref{fig:Temp_profile} show the presence of surface thermal inversion reminiscent of Earth polar night temperature profiles at almost all locations, except the substellar point. \cite{Zalucha2013} similarly report for their hot terrestrial GJ~1214b-like planet with $P_{rot}=1.58$~days global surface thermal inversion. However, although other studies also report surface thermal inversion, they appear to be more localized phenomena: \cite{Joshi1997} find them only at the anti-stellar point, and \cite{Edson2011} also find them locally for some rotation periods. In fact, they do not explicitly report them but they can be inferred from the potential temperature as shown in their Figure (4).

Both, \cite{Zalucha2013} and our model, use Rayleigh friction with $\tau_{fric}=1$~days, inspired by \textit{HS94}, whereas \cite{Joshi1997} and \cite{Edson2011} use Earth-like latent heat exchange and report surface thermal inversion at the nightside. Therefore, it appears reasonable that the differences are at least in part attributable to difference in surface-atmosphere interaction description. In principle, \cite{HengVogt2011} already showed that changing the friction timescales has a profound effect on the surface flow. However, first, they didn't address thermal inversion, second, they changed surface friction and radiative timescales simultaneously, third, their temperature retained too many Earth-centric assumptions and didn't properly address conditions at nightside.

Furthermore, we report additional thermal inversion at the top of the atmosphere for mid-latitudes in  $P_{rot}=10$~days model as shown in Figure~\ref{fig:Temp_profile}, left panel. This thermal inversion pattern is also very similar to \cite{Zalucha2013}, but, interestingly, it disagrees with the vertical temperature profiles for $P_{rot}=36.5$~days, where we have diminished thermal inversion on top.

\begin{figure*}
\includegraphics[width=0.495\textwidth]{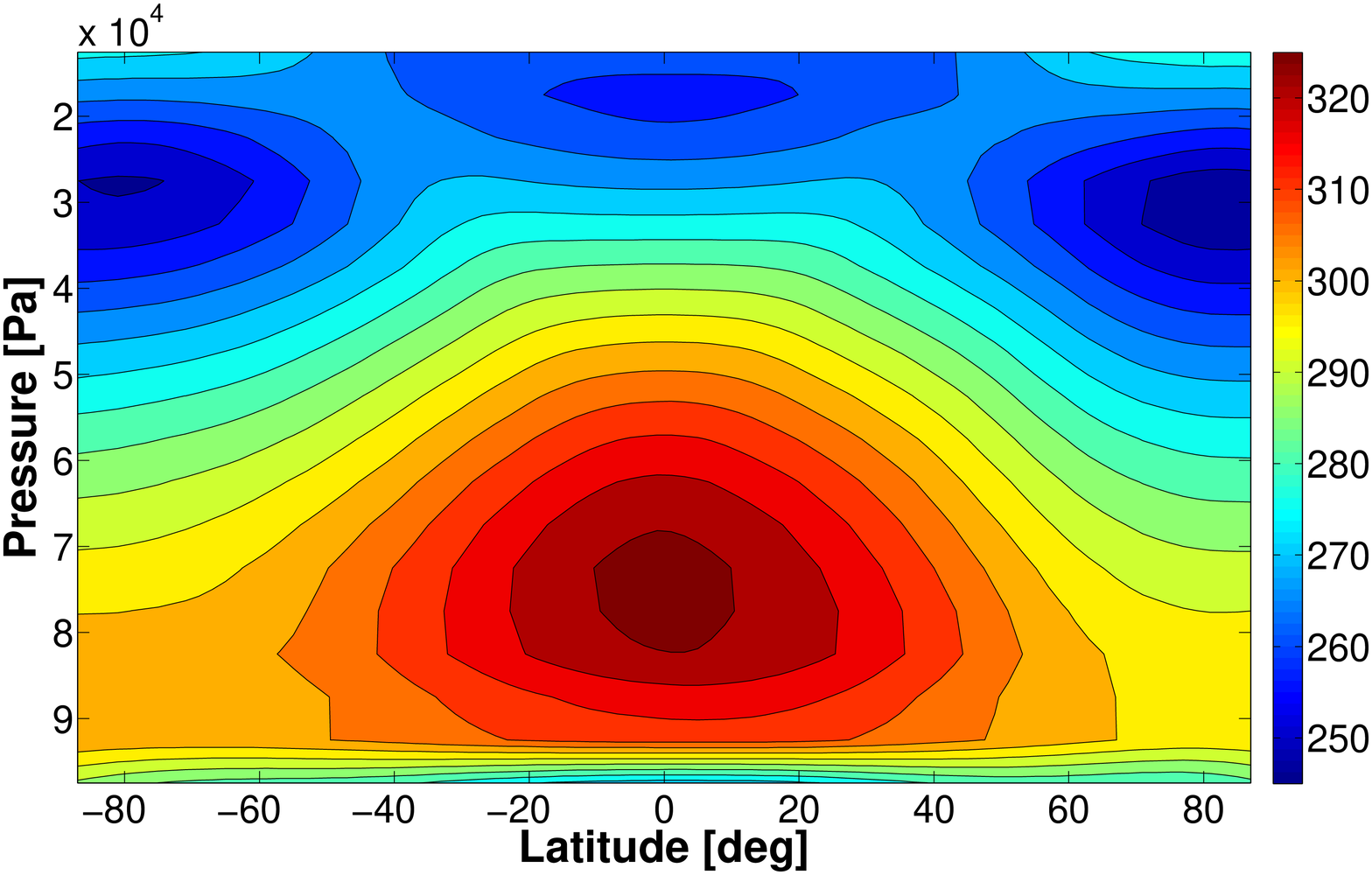}
\includegraphics[width=0.495\textwidth]{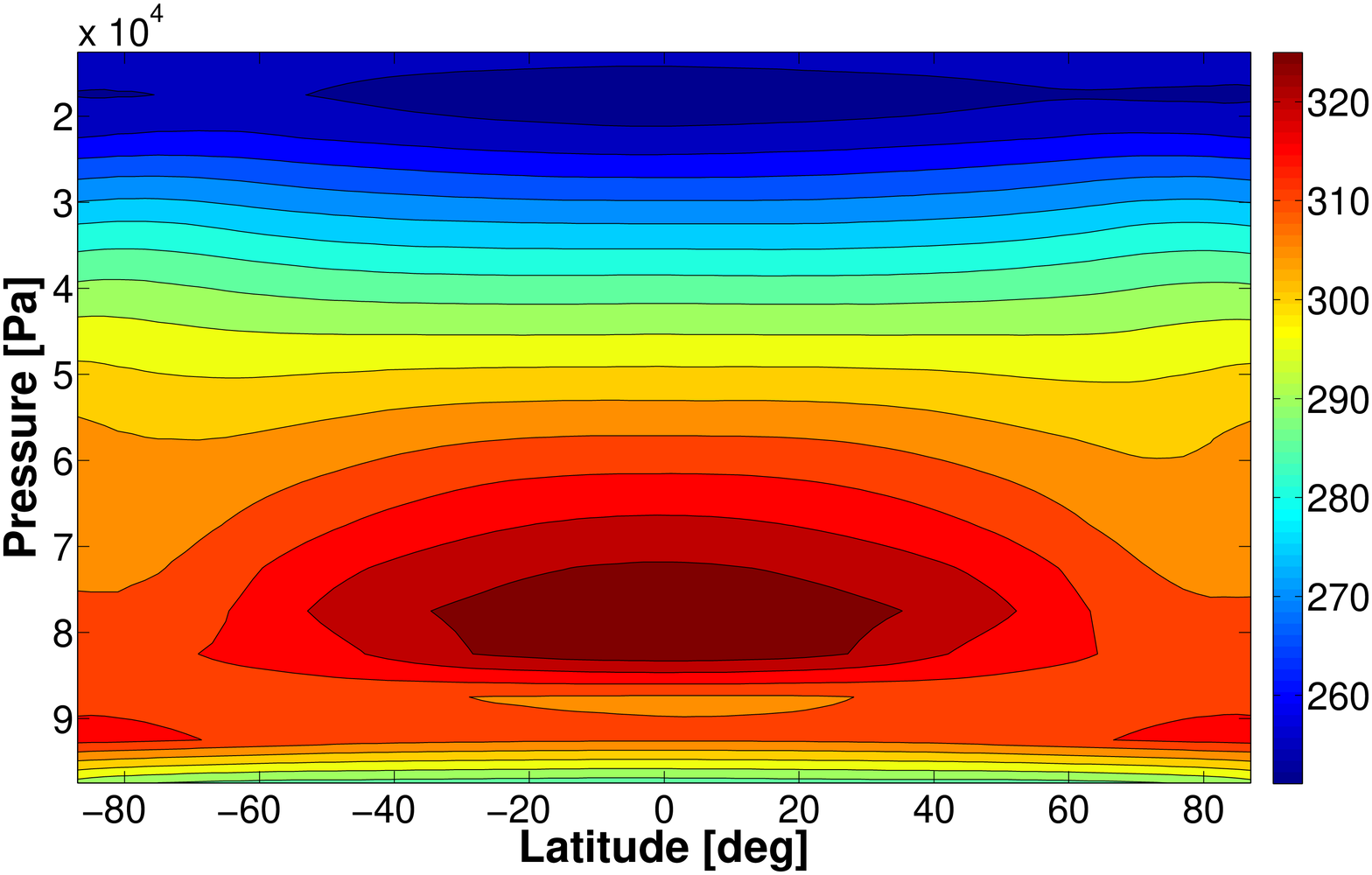}
\caption{Vertical cross section of the temperature at the antistellar side (LT=0 h) for $P_{rot}=10$~days (left panel) and for $P_{rot}=36.5$~days (right panel). The data are averaged over 1000 days. Contour levels are 5~K.}
\label{fig:Temp_profile_ap}
\end{figure*}

The connection between dynamics and the temperature profiles is more readily apparent when consulting the vertical slices of temperature at the anti-stellar nightside (Figure~\ref{fig:Temp_profile_ap}). Apparently, equatorial superrotating planets have a greater horizontal temperature gradient than atmospheres with more divergent dynamics and thus reduced equatorial superrotation, because the dynamics in superrotating atmospheres distributes heat predominantly along the equator. Consequently, the thermal inversions at the top of the atmosphere are stronger in atmospheres with faster rotation $P_{rot}$ period than compared to atmospheres with more divergent flow that distribute heat more evenly. Indeed, \cite{Edson2011} report that the slope of the potential temperature becomes less and less steep for slower rotation, which also corresponds to a decrease in horizontal temperature gradient, in agreement with our models. This confirms an adiabatic heating mechanism as the source of the upper atmosphere thermal inversion as the air in the upper atmosphere `falls' deeper at the cold nightside latitudes and is thus stronger heated than in the slow rotating case. Furthermore, it has been shown in Section~(\ref{sec: Horiz}) that the warm temperature anomalies at the top of the troposphere are correlated with strong horizontal temperature gradients in the middle troposphere.

\section{Conclusion}
\label{sec: Conclusion}

We present here a conceptional simple and versatile model for a tidally locked terrestrial planet by coupling a dry 3D-GCM with a Newtonian relaxation scheme using coherently derived radiative equilibrium temperatures and friction prescriptions with the same level of complexity used in \textit{HS94} for Earth.

We performed as proof of concept a small study for a tidally locked Super-Earth planet with Earth-like atmosphere and $P_{rot}=10$ and $P_{rot}=36.5$~days and found that already these two simulations show surprisingly different dynamics. The first case is still predominantly superrotating because of standing Kelvin and Rossby waves. The latter case shows on top of the atmosphere a divergent flow (\cite{Merlis2010}) with cyclonic and anti-cyclonic vortices embedded in a superrotating large amplitude planetary wave. Furthermore, it shows a decrease in zonal wind speed compared to the faster rotation model in agreement with \cite{Edson2011}. 

We furthermore report for our two `proof of concept' simulations, the emergence of zonal wind maxima that appear to be the result of an interplay between cyclonic vortices and the superrotating equatorial jet. It will be interesting to investigate how the zonal wind maxima change with increasing rotation period. Furthermore, it will be interesting to confirm that the vertical extent of such vortices can be changed when the surface friction timescale is varied.

We also report an upper atmosphere temperature inversion at the nightside mid-latitudes for faster rotation, also reported by \cite{Zalucha2013}. These diminish with increasing rotation period and thus decreasing horizontal temperature gradient suggesting adiabatic heating of the upwelling flow at the nightside-mid-latitudes.

Our study shows that dynamics is very efficient in transporting heat from the dayside towards the nightside, like \cite{Joshi1997} originally reported and has been reproduced in several studies since then (e,.g. \cite{Edson2011}, \cite{Joshi2003}). Indeed, the surface temperatures in our model generally allow for liquid water at the surface and in the $P_{rot}=36.5$ days-case even at the nightside. The planets investigated here can thus be considered habitable.

Comparison with other studies shows, however, that there is general disagreement between models in the details of heat transport: We report one circulation cell per hemisphere (like, \cite{Joshi1997}, \cite{Merlis2010}), whereas \cite{Edson2011} report two. Our nightside temperatures agree with the model of \cite{Joshi1997}, but are 100~K warmer than those of \cite{Edson2011} and \cite{Joshi2003}. Two obvious points that might explain these differences are temperature forcing at the nightside and surface friction assumptions.

Furthermore, we find near-global thermal inversion at the surface. But again, there seems to be disagreement about their prevalence. We find the best agreement with \cite{Zalucha2013}, who also use Rayleigh friction with $\tau_{fric}=1$~days, whereas \cite{Joshi1997} and \cite{Edson2011} use Earth-like latent heat exchange and report surface thermal inversion at the nightside. Therefore, it appears reasonable that the differences are at least in part attributable to difference in surface-atmosphere interaction description. Due to the conceptional simplicity of our model, the connection between surface friction and heat transport can be addressed easily by changing the surface friction timescales and extent of the planetary boundary layer. In a future study, we will explore variations between $\tau_{s,fric}=0.1 \to 100$~days and $p_{PBL}=(0.7\to 0.9)\times p_s$, taking into account the uncertainties from Solar System planets (Table~\ref{tab: PBL}), to investigate the influence of different surface friction scenarios on surface temperatures and circulation.

Interestingly, we find dynamics that hasn't been reported by other climate models. We identify Walker-like circulation along the equator between the anti-stellar point and the dawn-terminator in the mid-troposphere. We identify in the vertical velocities two such circulations in the meridional direction for $P_{rot}=10$~days and one for $P_{rot}=36.5$~days. These warrant in itself further investigation.

We propose, therefore, that our dry 3D GCM with simplified forcing is particularly valuable in the regime between strictly superrotating and strictly divergent atmospheres, e.g. for $3 \leq P_{rot}\leq 100$~days. Incidentally, these are the rotation periods that are expected for tidally locked planets in the habitable zone of M dwarf stars. We further conclude that our 3D model can extent the investigations of \cite{Showmanbook2013} by exploring the transit between superrotating and divergent flow regime not only for different thermal forcings but also under variation of Coriolis force. Last but not least, we can easily and comprehensibly change our model to also investigate different atmospheric compositions and surface pressures.

\section*{Acknowledgments}
We acknowledge support from the KU Leuven IDO project IDO/10/2013. The simulations were calculated at the VSC (flemish supercomputer center) funded by the Hercules foundation and the Flemish government. L.C. would further like to thank Adam Showman for useful discussions about fundamental climate dynamics and MITgcm properties during a research visit, Nikole A. Lewis for advise on the adaptation of MITgcm, Andras Zsom for useful discussions regarding the greenhouse model, and Olivia Venot for useful comments on the paper. We would also like to thank the anonymous referee for helpful comments and remarks.

\bibliography{GCM}

\begin{thebibliography}{}

\bibitem[\protect\citeauthoryear{{Adcroft}, {Campin}, {Hill} \&
  {Marshall}}{{Adcroft} et~al.}{2004}]{Adcroft2004}
{Adcroft} A.,  {Campin} J.-M.,  {Hill} C.,    {Marshall} J.,  2004, Monthly
  Weather Review, 132, 2845

\bibitem[\protect\citeauthoryear{{Bailey} \& {Kedziora-Chudczer}}{{Bailey} \&
  {Kedziora-Chudczer}}{2012}]{Bailey2012}
{Bailey} J.,  {Kedziora-Chudczer} L.,  2012, \mnras, 419, 1913

\bibitem[\protect\citeauthoryear{{Bordi}, {Fraedrich}, {Ghil} \&
  {Sutera}}{{Bordi} et~al.}{2009}]{Bordi2009}
{Bordi} I.,  {Fraedrich} K.,  {Ghil} M.,    {Sutera} A.,  2009, Journal of
  Atmospheric Sciences, 66, 1366

\bibitem[\protect\citeauthoryear{{Brown}, {Lebreton} \& {Waite}}{{Brown}
  et~al.}{2010}]{Brown2010}
{Brown} R.~H.,  {Lebreton} J.-P.,    {Waite} J.~H.,  2010, {Titan from
  Cassini-Huygens}

\bibitem[\protect\citeauthoryear{{Durran}}{{Durran}}{1991}]{Durran1991}
{Durran} D.~R.,  1991, Monthly Weather Review, 119, 702

\bibitem[\protect\citeauthoryear{{Edson}, {Lee}, {Bannon}, {Kasting} \&
  {Pollard}}{{Edson} et~al.}{2011}]{Edson2011}
{Edson} A.,  {Lee} S.,  {Bannon} P.,  {Kasting} J.~F.,    {Pollard} D.,  2011,
  \icarus, 212, 1

\bibitem[\protect\citeauthoryear{{Elachi}}{{Elachi}}{2005}]{Elachi2005}
{Elachi} C. e.~a.,  2005, Science, 308, 970

\bibitem[\protect\citeauthoryear{{Forget}, {Hourdin}, {Fournier}, {Hourdin},
  {Talagrand}, {Collins}, {Lewis}, {Read} \& {Huot}}{{Forget}
  et~al.}{1999}]{Forget1999}
{Forget} F.,  {Hourdin} F.,  {Fournier} R.,  {Hourdin} C.,  {Talagrand} O.,
  {Collins} M.,  {Lewis} S.~R.,  {Read} P.~L.,    {Huot} J.-P.,  1999, \jgr,
  104, 24155

\bibitem[\protect\citeauthoryear{{Gerber} \& {Vallis}}{{Gerber} \&
  {Vallis}}{2007}]{Gerber2007}
{Gerber} E.~P.,  {Vallis} G.~K.,  2007, Journal of Atmospheric Sciences, 64,
  3296

\bibitem[\protect\citeauthoryear{{Gierasch} \& {Goody}}{{Gierasch} \&
  {Goody}}{1972}]{Goody1972}
{Gierasch} P.~J.,  {Goody} R.~M.,  1972, Journal of Atmospheric Sciences, 29,
  400

\bibitem[\protect\citeauthoryear{{Grenfell}, {Gebauer}, {Godolt}, {Palczynski},
  {Rauer}, {Stock}, {von Paris}, {Lehmann} \& {Selsis}}{{Grenfell}
  et~al.}{2013}]{Grenfell2013}
{Grenfell} J.~L.,  {Gebauer} S.,  {Godolt} M.,  {Palczynski} K.,  {Rauer} H.,
  {Stock} J.,  {von Paris} P.,  {Lehmann} R.,    {Selsis} F.,  2013,
  Astrobiology, 13, 415

\bibitem[\protect\citeauthoryear{{Haberle}, {Houben}, {Barnes} \&
  {Young}}{{Haberle} et~al.}{1997}]{Haberle1997}
{Haberle} R.~M.,  {Houben} H.,  {Barnes} J.~R.,    {Young} R.~E.,  1997, \jgr,
  102, 9051

\bibitem[\protect\citeauthoryear{{Hanel}, {Conrath}, {Herath}, {Kunde} \&
  {Pirraglia}}{{Hanel} et~al.}{1981}]{Hanel1981}
{Hanel} R.,  {Conrath} B.,  {Herath} L.,  {Kunde} V.,    {Pirraglia} J.,  1981,
  \jgr, 86, 8705

\bibitem[\protect\citeauthoryear{{Hedelt}, {von Paris}, {Godolt}, {Gebauer},
  {Grenfell}, {Rauer}, {Schreier}, {Selsis} \& {Trautmann}}{{Hedelt}
  et~al.}{2013}]{Hedelt2013}
{Hedelt} P.,  {von Paris} P.,  {Godolt} M.,  {Gebauer} S.,  {Grenfell} J.~L.,
  {Rauer} H.,  {Schreier} F.,  {Selsis} F.,    {Trautmann} T.,  2013, \aap,
  553, A9

\bibitem[\protect\citeauthoryear{{Held}}{{Held}}{2005}]{Held2005}
{Held} I.~M.,  2005, Bulletin of the American Meteorological Society, 86, 1609

\bibitem[\protect\citeauthoryear{{Held} \& {Suarez}}{{Held} \&
  {Suarez}}{1994}]{HeldSuarez1994}
{Held} I.~M.,  {Suarez} M.~J.,  1994, Bulletin of the American Meteorological
  Society, 75, 1825

\bibitem[\protect\citeauthoryear{{Heng}, {Mendonca} \& {Lee}}{{Heng}
  et~al.}{2014}]{Heng2014}
{Heng} K.,  {Mendonca} J.,    {Lee} J.,  2014, ArXiv e-prints

\bibitem[\protect\citeauthoryear{{Heng} \& {Vogt}}{{Heng} \&
  {Vogt}}{2011}]{HengVogt2011}
{Heng} K.,  {Vogt} S.~S.,  2011, \mnras, 415, 2145

\bibitem[\protect\citeauthoryear{{Holton}}{{Holton}}{1992}]{Holton}
{Holton} J.~R.,  1992, {An introduction to dynamic meteorology}

\bibitem[\protect\citeauthoryear{{Hu}, {Seager} \& {Bains}}{{Hu}
  et~al.}{2012}]{Hu2012}
{Hu} R.,  {Seager} S.,    {Bains} W.,  2012, \apj, 761, 166

\bibitem[\protect\citeauthoryear{{Iro}, {B{\'e}zard} \& {Guillot}}{{Iro}
  et~al.}{2005}]{Iro2005}
{Iro} N.,  {B{\'e}zard} B.,    {Guillot} T.,  2005, \aap, 436, 719

\bibitem[\protect\citeauthoryear{{Irvine}}{{Irvine}}{1968}]{Irvine1968}
{Irvine} W.~M.,  1968, Journal of Atmospheric Sciences, 25, 610

\bibitem[\protect\citeauthoryear{Jablonowski \& Williamson}{Jablonowski \&
  Williamson}{2011}]{Jablo2011}
Jablonowski C.,  Williamson D.,  2011, in Lauritzen P.,  Jablonowski C.,
  Taylor M.,   Nair R.,  eds, Lecture Notes in Computational Science and
  Engineering, Vol.~80, Numerical Techniques for Global Atmospheric Models.
Springer Berlin Heidelberg, pp 381--493

\bibitem[\protect\citeauthoryear{{Joshi}}{{Joshi}}{2003}]{Joshi2003}
{Joshi} M.,  2003, Astrobiology, 3, 415

\bibitem[\protect\citeauthoryear{{Joshi}, {Haberle} \& {Reynolds}}{{Joshi}
  et~al.}{1997}]{Joshi1997}
{Joshi} M.~M.,  {Haberle} R.~M.,    {Reynolds} R.~T.,  1997, \icarus, 129, 450

\bibitem[\protect\citeauthoryear{{Joshi}, {Lewis}, {Read} \& {Catling}}{{Joshi}
  et~al.}{1995}]{Joshi1995}
{Joshi} M.~M.,  {Lewis} S.~R.,  {Read} P.~L.,    {Catling} D.~C.,  1995, \jgr,
  100, 5485

\bibitem[\protect\citeauthoryear{{Justus}, {Duvall} \& {Kller}}{{Justus}
  et~al.}{2004}]{Justus2004}
{Justus} C.~G.,  {Duvall} A.,    {Kller} V.~W.,  2004, in {Wilson} A.,  ed.,
  Planetary Probe Atmospheric Entry and Descent Trajectory Analysis and Science
  Vol.~544 of ESA Special Publication, {Engineering-level model atmospheres for
  Titan and Mars}.
pp 311--316

\bibitem[\protect\citeauthoryear{{Kataria}, {Showman}, {Fortney}, {Marley} \&
  {Freedman}}{{Kataria} et~al.}{2014}]{Kataria2014}
{Kataria} T.,  {Showman} A.~P.,  {Fortney} J.~J.,  {Marley} M.~S.,
  {Freedman} R.~S.,  2014, \apj, 785, 92

\bibitem[\protect\citeauthoryear{{Kliore} \& {Patel}}{{Kliore} \&
  {Patel}}{1980}]{Kliore1980}
{Kliore} A.~J.,  {Patel} I.~R.,  1980, \jgr, 85, 7957

\bibitem[\protect\citeauthoryear{{Kraucunas} \& {Hartmann}}{{Kraucunas} \&
  {Hartmann}}{2005}]{Kraucunas2005}
{Kraucunas} I.,  {Hartmann} D.~L.,  2005, Journal of Atmospheric Sciences, 62,
  371

\bibitem[\protect\citeauthoryear{{Lee}, {Lewis} \& {Read}}{{Lee}
  et~al.}{2007}]{Lee2007}
{Lee} C.,  {Lewis} S.~R.,    {Read} P.~L.,  2007, Journal of Geophysical
  Research (Planets), 112, 4

\bibitem[\protect\citeauthoryear{{Lewis}}{{Lewis}}{1971}]{Lewis1971}
{Lewis} J.~S.,  1971, Journal of Atmospheric Sciences, 28, 1084

\bibitem[\protect\citeauthoryear{{Lindal}, {Wood}, {Levy}, {Anderson},
  {Sweetnam}, {Hotz}, {Buckles}, {Holmes}, {Doms}, {Eshleman}, {Tyler} \&
  {Croft}}{{Lindal} et~al.}{1981}]{Lindal1981}
{Lindal} G.~F.,  {Wood} G.~E.,  {Levy} G.~S.,  {Anderson} J.~D.,  {Sweetnam}
  D.~N.,  {Hotz} H.~B.,  {Buckles} B.~J.,  {Holmes} D.~P.,  {Doms} P.~E.,
  {Eshleman} V.~R.,  {Tyler} G.~L.,    {Croft} T.~A.,  1981, \jgr, 86, 8721

\bibitem[\protect\citeauthoryear{{Lorenz} \& {Deweaver}}{{Lorenz} \&
  {Deweaver}}{2007}]{Lorenz2007}
{Lorenz} D.~J.,  {Deweaver} E.~T.,  2007, Journal of Geophysical Research
  (Atmospheres), 112, 10119

\bibitem[\protect\citeauthoryear{{Marshall}, {Adcroft}, {Campin}, {Hill} \&
  {White}}{{Marshall} et~al.}{2004}]{Marshall2004}
{Marshall} J.,  {Adcroft} A.,  {Campin} J.-M.,  {Hill} C.,    {White} A.,
  2004, Monthly Weather Review, 132, 2882

\bibitem[\protect\citeauthoryear{{Marshall}, {Hill}, {Perelman} \&
  {Adcroft}}{{Marshall} et~al.}{1997}]{Marshall1997}
{Marshall} J.,  {Hill} C.,  {Perelman} L.,    {Adcroft} A.,  1997, \jgr, 102,
  5733

\bibitem[\protect\citeauthoryear{{Marshall}, {Jones} \& {Hill}}{{Marshall}
  et~al.}{1998}]{Marshall1998}
{Marshall} J.,  {Jones} H.,    {Hill} C.,  1998, Journal of Marine Systems, 18,
  115

\bibitem[\protect\citeauthoryear{{Marshall} \& {Plumb}}{{Marshall} \&
  {Plumb}}{2008}]{Marshall}
{Marshall} J.,  {Plumb} R.,  2008, {Atmosphere, ocean, and climate dynamics}

\bibitem[\protect\citeauthoryear{{McKay}, {Lorenz} \& {Lunine}}{{McKay}
  et~al.}{1999}]{McKay1999}
{McKay} C.~P.,  {Lorenz} R.~D.,    {Lunine} J.~I.,  1999, \icarus, 137, 56

\bibitem[\protect\citeauthoryear{{McKay}, {Martin}, {Griffith} \&
  {Keller}}{{McKay} et~al.}{1997}]{McKay1997}
{McKay} C.~P.,  {Martin} S.~C.,  {Griffith} C.~A.,    {Keller} R.~M.,  1997,
  \icarus, 129, 498

\bibitem[\protect\citeauthoryear{{Menou}}{{Menou}}{2013}]{Menou2013}
{Menou} K.,  2013, \apj, 774, 51

\bibitem[\protect\citeauthoryear{{Merlis} \& {Schneider}}{{Merlis} \&
  {Schneider}}{2010}]{Merlis2010}
{Merlis} T.~M.,  {Schneider} T.,  2010, Journal of Advances in Modeling Earth
  Systems, 2, 13

\bibitem[\protect\citeauthoryear{{Molteni}}{{Molteni}}{2002}]{Molteni2002}
{Molteni} F.,  2002, Climate Dynamics, 20, 175

\bibitem[\protect\citeauthoryear{{Navarra} \& {Boccaletti}}{{Navarra} \&
  {Boccaletti}}{2002}]{Navarra2002}
{Navarra} A.,  {Boccaletti} G.,  2002, Climate Dynamics, 19, 467

\bibitem[\protect\citeauthoryear{{Pierrehumbert}}{{Pierrehumbert}}{2010}]{Pier%
rehumbert}
{Pierrehumbert} R.~T.,  2010, {Principles of Planetary Climate}

\bibitem[\protect\citeauthoryear{{Pollack} \& {Young}}{{Pollack} \&
  {Young}}{1975}]{Pollack1975}
{Pollack} J.~B.,  {Young} R.,  1975, Journal of Atmospheric Sciences, 32, 1025

\bibitem[\protect\citeauthoryear{{Polvani} \& {Kushner}}{{Polvani} \&
  {Kushner}}{2002}]{Polvani2002}
{Polvani} L.~M.,  {Kushner} P.~J.,  2002, \grl, 29, 1114

\bibitem[\protect\citeauthoryear{{Showman}, {Cooper}, {Fortney} \&
  {Marley}}{{Showman} et~al.}{2008}]{Showman2008}
{Showman} A.~P.,  {Cooper} C.~S.,  {Fortney} J.~J.,    {Marley} M.~S.,  2008,
  \apj, 682, 559

\bibitem[\protect\citeauthoryear{{Showman}, {Fortney}, {Lewis} \&
  {Shabram}}{{Showman} et~al.}{2013}]{Showman2013}
{Showman} A.~P.,  {Fortney} J.~J.,  {Lewis} N.~K.,    {Shabram} M.,  2013,
  \apj, 762, 24

\bibitem[\protect\citeauthoryear{{Showman}, {Fortney}, {Lian}, {Marley},
  {Freedman}, {Knutson} \& {Charbonneau}}{{Showman} et~al.}{2009}]{Showman2009}
{Showman} A.~P.,  {Fortney} J.~J.,  {Lian} Y.,  {Marley} M.~S.,  {Freedman}
  R.~S.,  {Knutson} H.~A.,    {Charbonneau} D.,  2009, \apj, 699, 564

\bibitem[\protect\citeauthoryear{{Showman} \& {Guillot}}{{Showman} \&
  {Guillot}}{2002}]{Showman2002}
{Showman} A.~P.,  {Guillot} T.,  2002, \aap, 385, 166

\bibitem[\protect\citeauthoryear{{Showman} \& {Polvani}}{{Showman} \&
  {Polvani}}{2011}]{Showman2011}
{Showman} A.~P.,  {Polvani} L.~M.,  2011, \apj, 738, 71

\bibitem[\protect\citeauthoryear{{Showman}, {Wordsworth}, {Merlis} \&
  {Kaspi}}{{Showman} et~al.}{2013}]{Showmanbook2013}
{Showman} A.~P.,  {Wordsworth} R.~D.,  {Merlis} T.~M.,    {Kaspi} Y.,  2013,
  {Atmospheric Circulation of Terrestrial Exoplanets}.
pp 277--326

\bibitem[\protect\citeauthoryear{{Toon}, {McKay}, {Ackerman} \&
  {Santhanam}}{{Toon} et~al.}{1989}]{Toon1989}
{Toon} O.~B.,  {McKay} C.~P.,  {Ackerman} T.~P.,    {Santhanam} K.,  1989,
  \jgr, 94, 16287

\bibitem[\protect\citeauthoryear{{Toon}, {Pollack}, {Ward}, {Burns} \&
  {Bilski}}{{Toon} et~al.}{1980}]{Toon1980}
{Toon} O.~B.,  {Pollack} J.~B.,  {Ward} W.,  {Burns} J.~A.,    {Bilski} K.,
  1980, \icarus, 44, 552

\bibitem[\protect\citeauthoryear{{Williamson}, {Olson} \&
  {Boville}}{{Williamson} et~al.}{1998}]{Will1998}
{Williamson} D.~L.,  {Olson} J.~G.,    {Boville} B.~A.,  1998, Monthly Weather
  Review, 126, 1001

\bibitem[\protect\citeauthoryear{{Zalucha}, {Michaels} \&
  {Madhusudhan}}{{Zalucha} et~al.}{2013}]{Zalucha2013}
{Zalucha} A.~M.,  {Michaels} T.~I.,    {Madhusudhan} N.,  2013, \icarus, 226,
  1743

\end{thebibliography}

\label{lastpage}
\end{document}